\documentclass[sigconf]{acmart}
\AtBeginDocument{%
  \providecommand\BibTeX{{%
    \normalfont B\kern-0.5em{\scshape i\kern-0.25em b}\kern-0.8em\TeX}}}

\copyrightyear{2022} 
\acmYear{2022} 
\setcopyright{acmcopyright}\acmConference[CIKM '22]{Proceedings of the 31st ACM International Conference on Information and Knowledge Management}{October 17--21, 2022}{Atlanta, GA, USA}
\acmBooktitle{Proceedings of the 31st ACM International Conference on Information and Knowledge Management (CIKM '22), October 17--21, 2022, Atlanta, GA, USA}
\acmPrice{15.00}
\acmDOI{10.1145/3511808.3557072}
\acmISBN{978-1-4503-9236-5/22/10}

\usepackage{ulem} 
\usepackage{enumitem}
\usepackage{threeparttable}
\usepackage{ulem}
\usepackage{booktabs}
\usepackage{multirow}
\usepackage{amsmath}
\usepackage{mathrsfs}
\usepackage{graphicx}
\usepackage[T1]{fontenc}
\usepackage{aecompl}
\usepackage[inkscapelatex=false]{svg}
\usepackage{subfigure}

\begin{document}

\title{IntTower: the Next Generation of  Two-Tower Model for Pre-Ranking System}




\author{Xiangyang Li}
\authornote{Co-first authors with equal contributions. Work done when Xiangyang Li was intern at Huawei Noah’s Ark Lab.}
\affiliation{%
  \institution{Peking University}
   \country{China}
  }
  \email{xiangyangli@pku.edu.cn}

 \author{Bo Chen}
 \authornotemark[1]
\affiliation{%
  \institution{Huawei Noah's Ark Lab}
\country{China}
}
  \email{chenbo116@huawei.com}
  
 \author{HuiFeng Guo}
 \authornote{Corresponding authors.}
\affiliation{%
  \institution{Huawei Noah's Ark Lab}
\country{China}
}
\email{huifeng.guo@huawei.com}

 \author{Jingjie Li}
\affiliation{%
  \institution{Huawei Noah's Ark Lab}
\country{China}
}
\email{lijingjie1@huawei.com}

 \author{Chenxu Zhu}
\affiliation{%
  \institution{Shanghai JiaoTong University}
\country{China}
}
\email{zhuchenxv@sjtu.edu.cn}

 \author{Xiang Long}
\affiliation{%
  \institution{Beijing University of Posts and Telecommunication}
\country{China}
}
\email{xianglong@bupt.edu.cn}

 \author{Sujian Li}
\affiliation{%
  \institution{Peking University}
\country{China}
}
\email{lisujian@pku.edu.cn}

 \author{Yichao Wang}
\affiliation{%
  \institution{Huawei Noah's Ark Lab}
\country{China}
}
\email{wangyichao5@huawei.com}

 \author{Wei Guo}
\affiliation{%
  \institution{Huawei Noah's Ark Lab}
\country{China}
}
\email{guowei67@huawei.com}

 \author{Longxia Mao}
\affiliation{%
  \institution{Huawei Technologies Co Ltd}
\country{China}
}
\email{maolongxia@huawei.com}

 \author{Jinxing Liu}
\affiliation{%
  \institution{Huawei Technologies Co Ltd}
\country{China}
}
\email{liujinxing5@huawei.com}

 \author{Zhenhua Dong}
\affiliation{%
  \institution{Huawei Noah's Ark Lab}
\country{China}
}
\email{dongzhenhua@huawei.com}

 \author{Ruiming Tang}
  \authornotemark[2]
\affiliation{%
  \institution{Huawei Noah's Ark Lab}
\country{China}
}
\email{tangruiming@huawei.com}

\renewcommand{\shortauthors}{Xiangyang Li et al.}
\begin{abstract}
Scoring a large number of candidates precisely in several milliseconds is vital for industrial pre-ranking systems. 
Existing pre-ranking systems primarily adopt the \textbf{two-tower} model since the ``user-item decoupling architecture'' paradigm is able to balance the \textit{efficiency} and \textit{effectiveness}.  However, the cost of high efficiency is the neglect of the potential information interaction between user and item towers, hindering the prediction accuracy critically. In this paper, we show it is possible to design a two-tower model that  emphasizes both information interactions and inference efficiency. The proposed model, IntTower (short for \textit{Interaction enhanced Two-Tower}), consists of Light-SE, FE-Block and CIR modules. 
Specifically, lightweight Light-SE module is used to identify the importance of different features and obtain refined feature representations in each tower. FE-Block module performs fine-grained and early feature interactions to capture the interactive signals between user and item towers explicitly and CIR module leverages a contrastive interaction regularization to further enhance the interactions implicitly.
Experimental results on three public datasets show that IntTower outperforms the SOTA pre-ranking models significantly and even achieves comparable performance in comparison with the ranking models.
Moreover, we further verify the effectiveness of IntTower on a large-scale advertisement pre-ranking system. 
The code of IntTower is publicly available\footnote{https://github.com/archersama/IntTower}.

\end{abstract}




\maketitle

\section{Introduction}

\begin{figure}[htbp]
\centering
\includegraphics[scale=0.24]{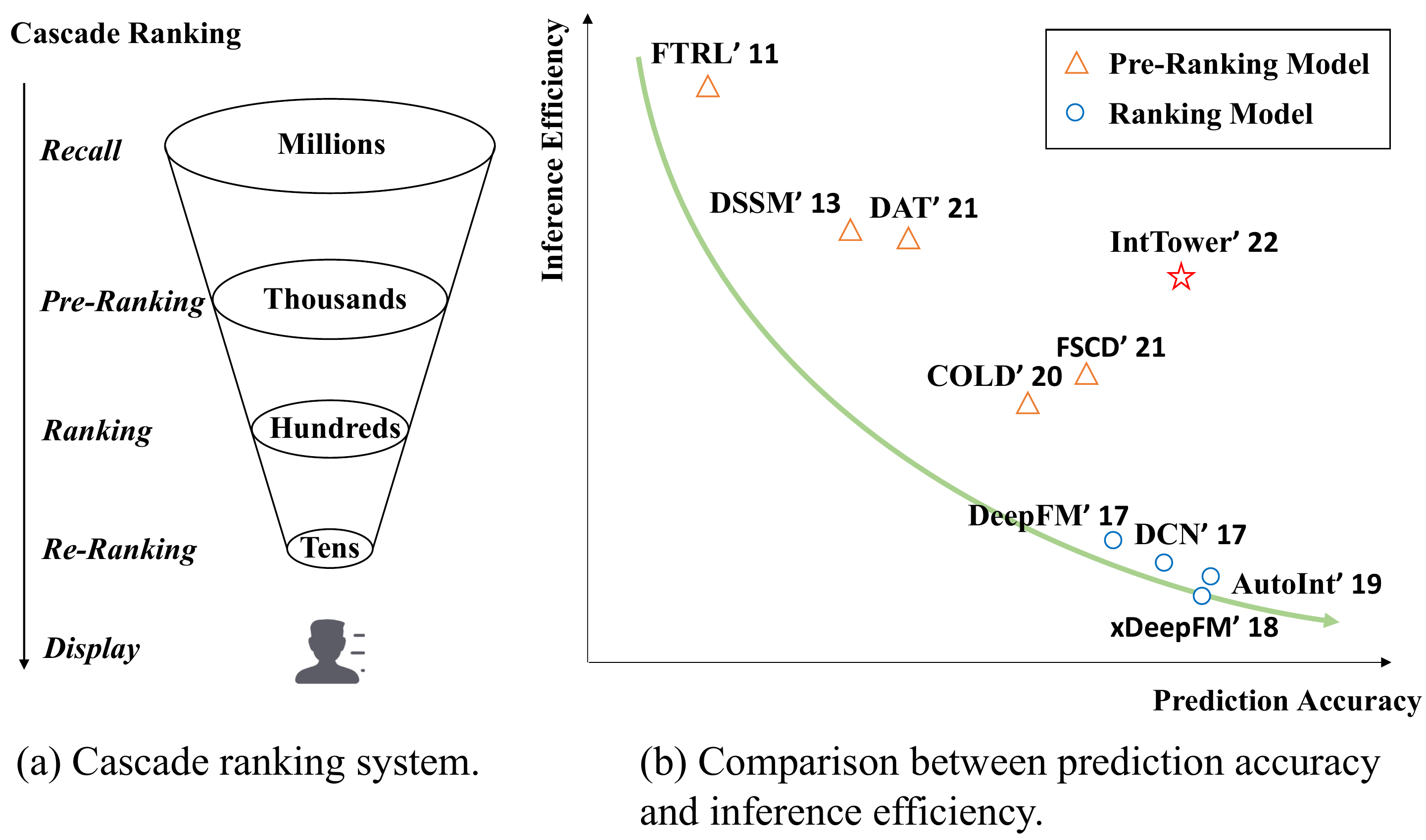}
\caption{\small{The multi-stage cascade ranking architecture and the comparison of model prediction accuracy and inference efficiency.}}
\label{ee}
\vspace{-0.3cm}
\end{figure}

Existing industrial information services, such as recommender system, search engine, and advertisement system, are multi-stage cascade ranking architecture, which contributes to balancing the efficiency and effectiveness in comparison with the single-stage architecture~\cite{rankflow}. Typical cascade ranking system consists of Recall, Pre-Ranking, Ranking, and Re-Ranking stages (Figure~\ref{ee}(a)).
The early stages face a massive number of candidates, and thus using simple models (\textit{e.g.}, LR and DSSM~\cite{dssm}) to guarantee low inference latency. On the contrary, the later stages pursue subtly selected items that meet the user's preferences, and hence complex models (\textit{e.g.}, DeepFM~\cite{guo2017deepfm} and AutoInt~\cite{autoint}) are conducive to improve the prediction accuracy. 

Pre-ranking stage is in the middle of a cascade link, which is absorbed in preliminarily filtering items (thousands of scale) retrieved from the previous recall stage and generating candidates (hundreds of scale) for the subsequent ranking stage. Therefore, both effectiveness and efficiency need to be carefully considered.
Figure~\ref{ee}(b) depicts some representative pre-ranking and ranking models from the perspective of \textit{prediction accuracy} and \textit{inference efficiency}.
Compared with the ranking models, pre-ranking models need to score more candidate items for each user request. Therefore, pre-ranking models have higher inference efficiency while weaker prediction performance due to simpler structure.

In the evolution of pre-ranking system, LR (Logistic Regression)~\cite{mcmahan2013ad} is the most basic personalized pre-ranking model, which is widely used in the shallow machine learning era~\cite{log1,log2,log3}. 
With the rise of deep learning, many industrial companies deploy various deep models in their commercial systems. 
The dominant pre-ranking model in industry is two-tower model~\cite{dssm} (\textit{i.e.}, DSSM), which utilizes neural networks to capture the interactive signals within the user/item towers. 
Moreover, the item representations can be pre-calculated offline and stored in the fast retrieval container. During the online serving, only user representations are required to be calculated in real time while the representations of candidate items can be retrieved directly.
These ``\textit{user-item decoupling architecture}'' paradigm provides sterling efficiency. 
Besides, COLD~\cite{cold} and FSCD~\cite{fscd} propose a single-tower structure to fully model feature interaction and further improve the prediction accuracy.

Despite great promise, existing pre-ranking models are difficult to balance model effectiveness and inference efficiency. 
For the two-tower model, the cost of high efficiency is the neglect of the information interaction between user and item towers. Two towers perform intra-tower information extraction parallelly and independently, and the learned latent representations do not interact until the output layer, which is referred to as ``\textit{Late Interaction}''~\cite{colbert}, hindering the model performance critically. However, the interactive signals between user features and item features are vital for prediction~\cite{dcn}.
Though DAT~\cite{dat} attempts to alleviate this issue by implicitly modeling the information interaction between the two towers, the performance gain is still limited. 
As for the single-tower structure pre-ranking models (\textit{i.e.}, COLD and FSCD), although several optimization tricks are introduced for acceleration, the efficiency degradation is still severe ($\times$10).

To solve the efficiency-accuracy dilemma, we propose a next generation of two-tower model for  pre-ranking system, named \textit{\textbf{Int}eraction enhanced Two-\textbf{Tower}} (\textbf{IntTower}), as illustrated in the Figure~\ref{IntTower}. The core idea is to enhance the information interaction between user and item towers while keeping the ``\textit{user-item decoupling architecture}'' paradigm. By introducing fine-grained feature interaction modeling, the model capacity of two-tower can be improved significantly and the sterling inference efficiency can be maintained.
Specifically, IntTower first leverages a lightweight \textit{Light-SE} module to identify the importance of different features and obtain refined feature representations. 
Based on the refined representations, user and item towers leverage multi-layer nonlinear transformation to extract latent representations.
To capture the interactive signals between user and item representations, IntTower designs \textit{FE-Block} module and \textit{CIR} module from \textbf{explicit} and \textbf{implicit} perspectives, respectively.
FE-Block module performs fine-grained and early feature interactions between multi-layer user representations and last-layer item representation. Thus, multi-level feature interaction modeling contributes to improving the prediction accuracy while the user-item decoupling architecture enables high inference efficiency.
Moreover, CIR proposes a contrastive interaction regularization to further enhance the interactions between user and item representations.

Our main contributions are summarized as follows: (1) We propose IntTower, the next generation of two-tower model for the pre-ranking system, which emphasizes both high prediction accuracy and inference efficiency. (2) IntTower leverages a lightweight Light-SE module to obtain refined feature representations. Based on this, FE-Block module and CIR module are proposed to capture the interactive signals between user and item representations from explicit and implicit perspectives. (3) Comprehensive experiments are conducted on three public datasets to demonstrate the superiority of IntTower over prediction accuracy and inference efficiency. Moreover, we further verify the effectiveness of IntTower on a large-scale advertisement pre-ranking system.

\section{RELATED WORK}
\subsection{Pre-Ranking System}
 Pre-ranking system is located in middle stage of cascade ranking system and needs to predict thousands of candidate items in a few milliseconds, which is sensitive to both model effectiveness and efficiency. 
LR model is the simplest personalized pre-ranking model, which has strong fitting capability and is widely used in the shallow machine learning era. Besides, FM model~\cite{fm} is also popular for the pre-ranking stage, which models the low-order feature interactions using factorized parameters and can be calculated in linear time.

With the rise of deep learning~\cite{explore,low_resource}, neural network-based models are gradually introduced into industrial pre-ranking systems. 
The most important one is two-tower model~\cite{dssm} (\textit{i.e.}, DSSM), which designs two-tower to capture the interactive signals within the user/item towers. Moreover, the ``\textit{user-item decoupling architecture}'' designing paradigm enables sterling inference efficiency. Due to the pre-storage of the item representations, during the online serving, only a single inference is required to obtain user representation for each request. However, the interaction modeling is inadequate for DSSM, hindering the model performance critically.
To alleviate this issue, DAT~\cite{dat} implicitly models the information interaction between the two towers with an Adaptive-Mimic Mechanism. Compared with the DSSM and DAT, our proposed IntTower proposes FE-Block and CIR modules to capture the interactive signals between two-tower from explicit and implicit perspectives, respectively.

To improve the prediction accuracy, COLD~\cite{cold} and FSCD~\cite{fscd} leverage the single-tower structure to fully model feature interaction. To balance the model efficiency and effectiveness, they adopt optimization tricks (\textit{e.g.}, parallel computation and semi-precision calculation) and complexity-aware feature selection for efficiency acceleration, respectively. However, the single-tower structure determines the need for multiple inferences for each user request given multiple candidate items, which is more time-consuming than two-tower structure. 

\subsection{Ranking System}
For ranking system, whose scale of predicted candidate items is much smaller than that in pre-ranking, user preferences over items need to be learned more accurately. Therefore, models deployed in the ranking system focus on extracting feature interactions with various operations. DNN model is widely used to capture the high-order implicit feature interactions; while the explicit feature interaction modeling is diverse for different models. 
Wide\&Deep~\cite{widedeep} utilizes handcrafted cross features to memorize important patterns. PNN~\cite{pnn} and DeepFM~\cite{guo2017deepfm} use the inner product to capture pair-wise interactions. CFM~\cite{cfm} and FGCNN~\cite{fgcnn} leverages the convolution operation to identify the local patterns and models feature interactions. DCN~\cite{dcn} and EDCN~\cite{edcn} use Cross Network to learn certain bounded-degree feature interactions, while xDeepFM~\cite{xdeepfm} extends to vector-wise level with a Compressed Interaction Network. AutoInt~\cite{autoint} and DIN~\cite{din} leverage the Attention Network to model high-order feature interactions and user historical behaviors.
However, these ranking models belong to single-tower structure, which have large serving latency and cannot be deployed in the pre-ranking system directly.
Instead, our IntTower performs fine-grained and early interaction modeling with two-tower structure, achieving comparable prediction accuracy while higher inference efficiency than the ranking models.

\section{Preliminary}
\label{sect:preliminary}
\begin{figure}[htbp]
\centering
\includegraphics[scale=0.3]{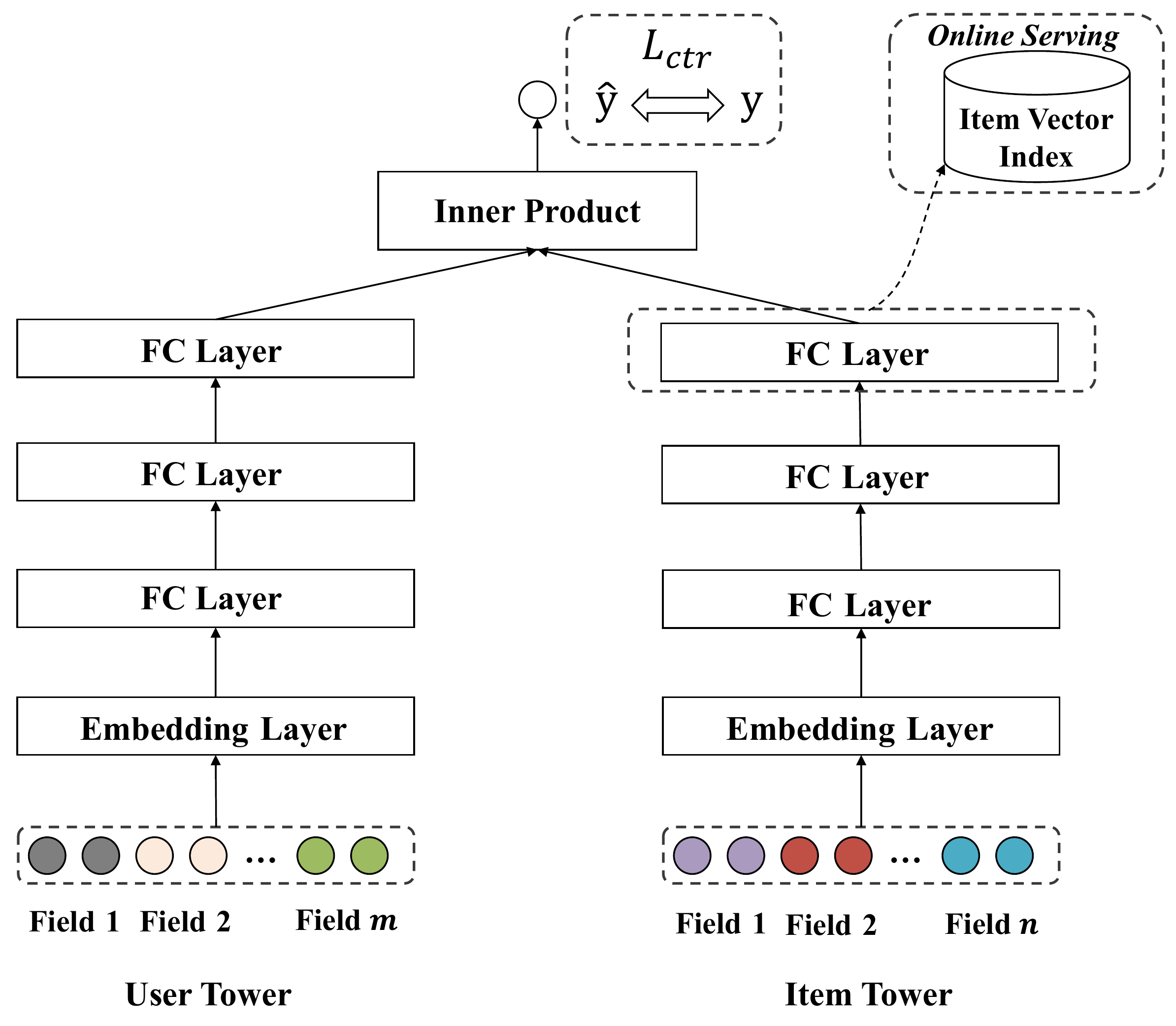}
\caption{\small{The overview of neural network based two-tower model.}}
\label{dssm}
\end{figure}

The neural network based two-tower, one of the state-of-the-art pre-ranking models, has excellent trade-off between prediction accuracy and inference efficiency. In this section, we first present the details of two-tower, then demonstrate both advantages and disadvantages of applying two-tower model in the pre-ranking system. The architecture is shown in Figure~\ref{dssm}, which consists of two parallel sub-networks.

Specifically, the dataset for training pre-ranking models consists of instances $(\mathbf{x}, y)$, where $y \in \{0,1\}$ indicates the user-item feedback label~\cite{richardson2007predicting} that means positive signal when equals 1 and negative when it is 0,  features $\mathbf{x}$ can be divided into $m$ user-related features and $n$ item-related features, \textit{i.e.}, $\mathbf{x}=[\underbrace{x_1,x_2,\dots,x_m}_{\text{user-related}};\underbrace{x_1,x_2,\dots,x_n}_{\text{item-related}}]$.
Then these user-related features and item-related features are fed into the corresponding parallel sub-towers (\textit{i.e.}, user tower and item tower) to obtain user and item representations.
Take the user tower as an example.
The user-related features are first fed into the embedding layer to obtain user feature embeddings $\mathbf{e} = [\mathbf{e}_{1}, \mathbf{e}_{2}, ..., \mathbf{e}_{m}]$ via embedding look-up operation~\cite{zhang2016deep}, where $\mathbf{e}_{i}\in \mathbb{R}^{d}$ denotes the embedding of the $i$-th feature and $d$ is the embedding dimension. Then the feature embeddings are further processed by multiple FC (fully connected) layers:
\begin{equation}
\label{e1}
\begin{aligned}
\mathbf{h}^{i+1}=relu(\mathbf{W}^{i}\mathbf{h}^{i}+\mathbf{b}^{i}),
\end{aligned}
\end{equation}
where $\mathbf{W}^{i}\in \mathbb{R}^{d_{i+1}\times d_{i} }$, $\mathbf{b}^{i}\in \mathbb{R}^{d_{i} }$ are the weight and bias of the $i$-th FC layer,  and the $\mathbf{h}^{0}=\mathbf{e}$, $d_i$ is the width of the $i$-th FC layer and $d_0 = d\times m$. Finally, the user representation can be obtained by: $\mathbf{h}_u = L2Norm(\mathbf{h}^{L})$, where $L$ is the depth of FC layers. Similarly, the item representation $\mathbf{h}_v$ can learned via the item tower.
Finally, the output of the two-tower model $\hat{y}$ is the inner product of the user representation and item representation:
 \begin{equation}
\label{inner_product}
\begin{aligned}
\hat{y} =\mathbf{h}_u^{T}\mathbf{h}_v.
\end{aligned}
\end{equation}







Obviously, the two-tower model adopts ``user-item decoupling architecture'' designing paradigm, which is more efficient in terms of inference latency and computational resources.
The item representations $\mathbf{h}_i$ can be periodically pre-calculated offline and stored in the fast retrieval container.
During the online serving stage, for each user request, the user representation can be obtained by one online inference, and the $k$ candidate item representations can be gathered from the retrieval container for prediction (the scale of $k$ is thousands). Therefore, the overall time complexity is $O(N+kM)$, which is smaller than the single-tower structure with $O(kN)$ complexity, where $N$ is the cost of a neural network inference and $M$ is the cost of a vector retrieval and score ($N \gg M$).
However, the disadvantage of two-tower model is also apparent. The user/item representations do not interact until the prediction layer, which is referred to ``Late Interaction'', hindering the
model performance critically.
To enhance the performance of two-tower model, we propose IntTower, which provides fine-grained and early interaction modeling and maintains excellent efficiency.



\section{IntTower: Interaction enhanced Two-Tower  }
In this section, we will give  a detailed description of the proposed pre-ranking model IntTower. The core idea behind IntTower is to utilize fine-grained and early feature interactions to improve the performance of the two-tower model. Remarkably, IntTower considers the trade-off between prediction accuracy and inference efficiency, which is vital for the pre-ranking system. 

\begin{figure*}[htbp]
\centering
\includegraphics[scale=0.35]{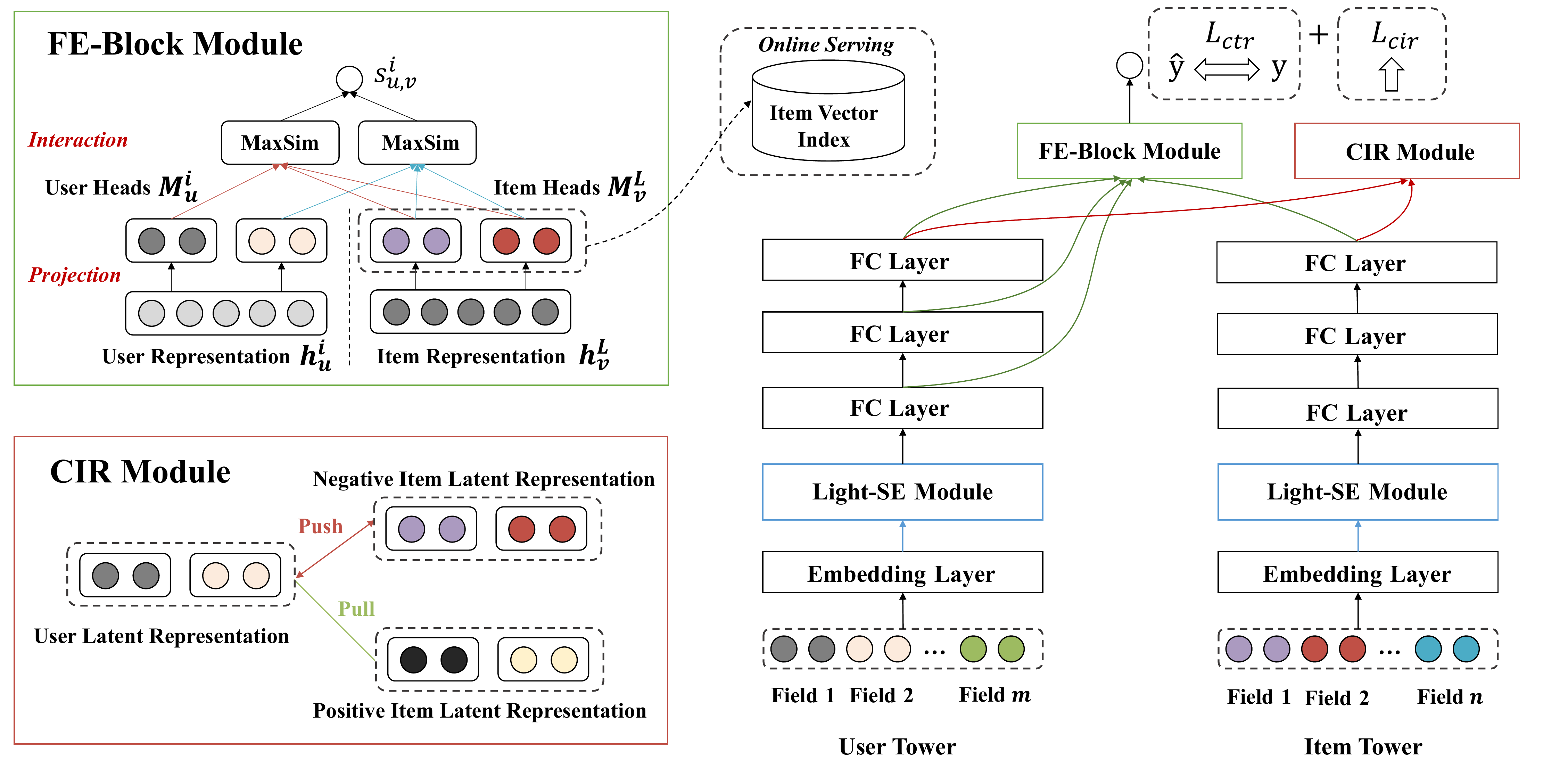}
\caption{\small{The architecture of IntTower.}}
\label{IntTower}

\end{figure*}

Figure~\ref{IntTower} depicts the overall architecture of IntTower. 
Based on the original Two-Tower model, IntTower introduces three module: \emph{Light-SE} (Lightweight SENET), \emph{FE Block} (Fine-grained and Early Feature Interaction Block), and \emph{CIR} (Contrastive Interaction Regularization).
Firstly, for both user tower and item tower, a lightweight Light-SE module is leveraged to identify the importance of different features and obtain refined feature representations. 
Then, FE-Block module and CIR module are proposed to capture the interactive signals between user and item representations, from the \textbf{explicit} and \textbf{implicit} perspectives, respectively.
Specifically, FE-Block performs fine-grained feature interactions between user and item representation explicitly, and CIR module implements a contrastive interaction regularization to further enhance the interactions implicitly.

\subsection{Light Feature Attention Mechanism}

\begin{figure}
	\centering
	\setlength{\belowcaptionskip}{-0.3cm}
	\setlength{\abovecaptionskip}{0cm}
	\subfigure[\small{Light-SE}]{
		\label{fig:lightse} 
		\includegraphics[height=110pt]{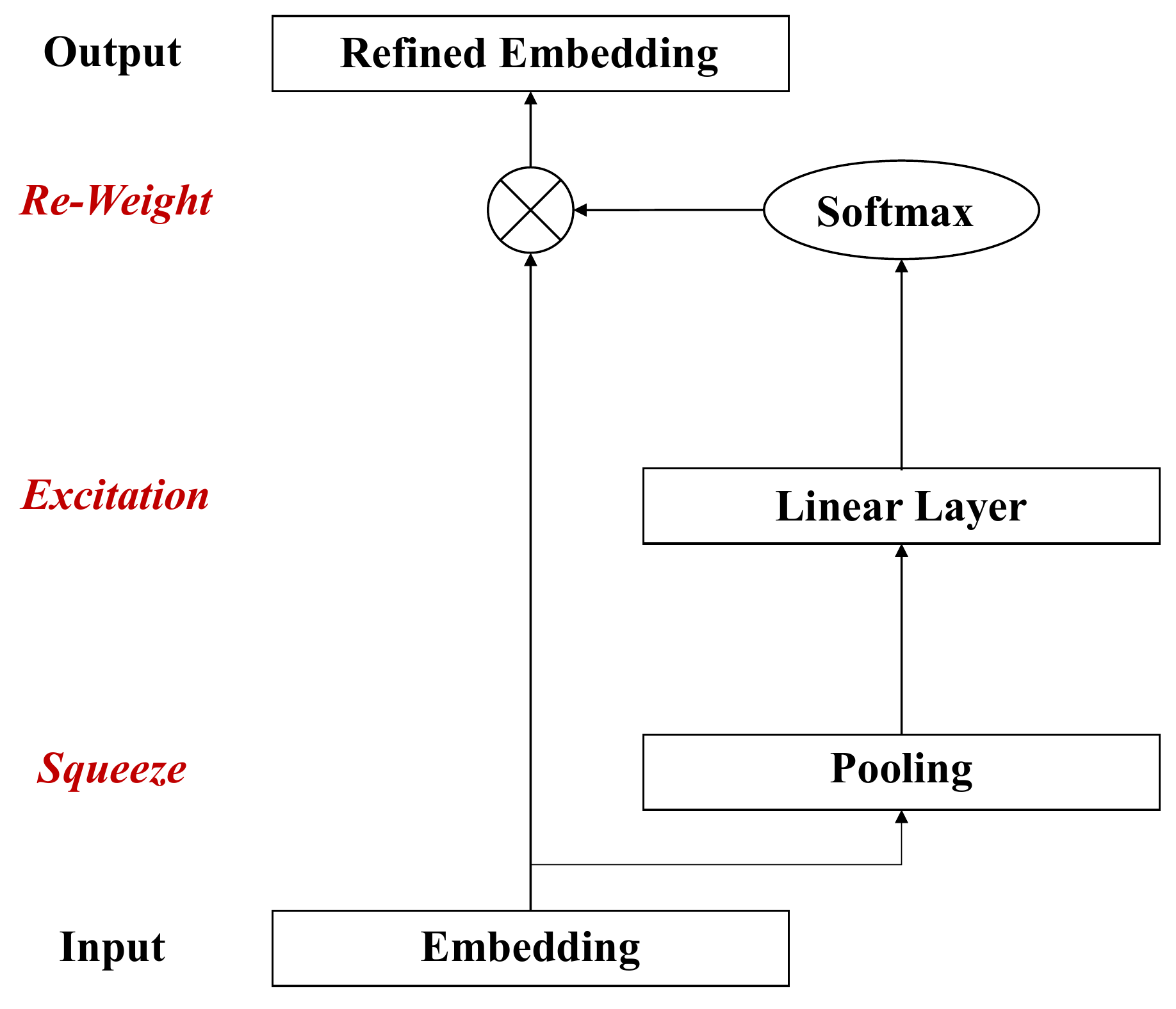}}
	\subfigure[\small{SENET}]{
		\label{fig:senet} 
		\includegraphics[height=110pt]{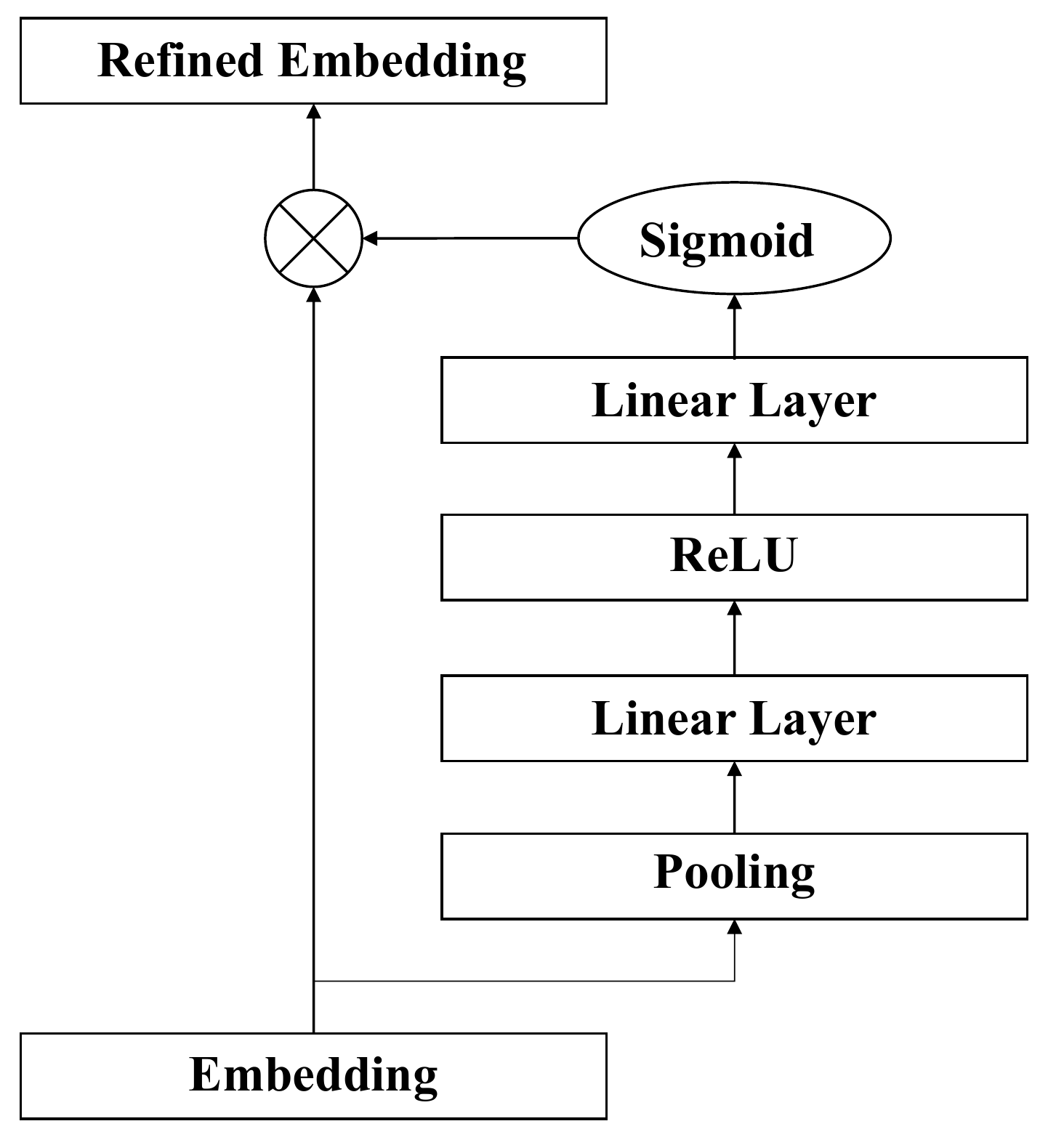}}
	\caption{\small{The comparison of Light-SE and SENET.}}
	\vspace{-0.2cm}
	\label{fig:lightse_senet}
\end{figure}

For the recommender system, the importance of different features for prediction is diverse. For example, in the e-commerce scenario, it is apparent that \texttt{user consumption level} is more important than \texttt{gender}.
Therefore, it is essential to automatically assign weights for different features to obtain refined feature representations, thus facilitating the subsequent feature interactions.
Some previous works, like COLD~\cite{cold} and FiBiNET~\cite{huang2019fibinet}, use SENET (Squeeze-and-Excitation Net)~\cite{hu2018squeeze} to identify the feature importance. 
However, for pre-ranking models, introducing SENET will undoubtedly increase the online serving latency and aggravate the system pressure. Thus, We propose a lightweight version of SENET, named Light-SE with fewer parameters and lower latency. 
As shown in Figure~\ref{fig:lightse_senet}, like SENET, Light-SE also consists of three steps: \textit{Squeeze}, \textit{Excitation} and \textit{Reweighing} step. 
\\\textbf{Squeeze.} This step is used to extract the ``statistical  information'' of  feature embedding. Specifically, we use the mean pooling operation to squeeze original user/item embeddings $\mathbf{e} = [\mathbf{e}_{1}, \mathbf{e}_{2} , \dots, \mathbf{e}_{m}] $ into vector $\mathbf{z} = [z_{1}, z_{2} , \dots, z_{m} ]$, where $z_{i}$ is a scalar value that reflects the global statistic information of the $i$-th user feature, shown as:
 \begin{equation}
\label{e1}
\begin{aligned}
z_{i} = f_{sq}(\mathbf{e}_{i}) =  \frac{1}{d}\sum_{t=1}^{d}\mathbf{e}_{i}^{d}.
\end{aligned}
\end{equation}
\\\textbf{Excitation and Re-Weight}. The next step is to learn the importance of each feature based on the statistical vector $\mathbf{z}$. Previous works~\cite{cao2019gcnet,xue2019danet} has shown that using two FC layers of SENET is not necessary. Thus, here we use a more lightweight approach with a single FC layer to obtain the feature importance. Besides, we replace the Sigmoid with Softmax function to take the relative relationship of different feature importances into consideration comprehensively.
Formally, the weight vector $\mathbf{k}\in \mathbb{R}^{m}$ can be calculated as
follows:

 \begin{equation}
\label{e1}
\begin{aligned}
\mathbf{k} = f_{ex}(\mathbf{z}) = softmax(\mathbf{W}\mathbf{z}+\mathbf{b}),
\end{aligned}
\end{equation}
where $\mathbf{W}\in \mathbb{R}^{m\times m}$ and $\mathbf{b}\in \mathbb{R}^{m}$ are the transform matrix and bias vector, respectively.
The last step is the re-weight step, which does the multiplication between the original feature embedding $\mathbf{e}$ and the weight vector $\mathbf{k}$, and outputs the refined user feature embedding $\tilde{\mathbf{e}}$, which can be calculated as follows:
 \begin{equation}
\label{e1}
\begin{aligned}
\tilde{\mathbf{e}} = f_{rw}(\mathbf{k}, \mathbf{e}) = [k_{1}\cdot \mathbf{e}_{1}, k_{2}\cdot \mathbf{e}_{2}, \dots , k_{m}\cdot \mathbf{e}_{m}].
\end{aligned}
\end{equation}

Based on the refined user and item feature representations, the user and item tower can better capture feature interactions.

\subsection{Fine-grained and Early Feature  Interaction}
As mentioned in Section~\ref{sect:preliminary}, the two-tower model performs ``\textit{Late Interaction}'' modeling and the user/item representations do not interact until the prediction layer (E.q.(\ref{inner_product})). This approach reduces the online inference latency, yet makes compromise on accuracy. However, effectively modeling feature interactions contributes to improving prediction accuracy~\cite{zhang2021deep,edcn}. 
Inspired by the ColBERT~\cite{colbert}, we propose FE-Block with ``\textit{Early Interaction}'', which improves the performance of the two-tower model with fine-grained interactions and is friendly enough for computing power costs.
FE-Block consists of multiple layers of interaction modeling, each of which contains two steps: \textit{Projection} and \textit{Interaction} step.
It's worth noting that, in order to keep the ``user-item decoupling architecture'' and save a single vector for each item thus accelerating the online inference procedure, FE-Block performs fine-grained feature interactions between multi-layer user representations and last-layer item representation.
\\\textbf{Projection.} This step is to map the user/item representations into the latent space and extract more informative knowledge.
As Figure~\ref{IntTower} shows, FE-Block extracts the latent information from each FC layer for user tower, while the last FC layer only for item tower.
For user/item tower, each representation is transformed into $H$ head sub-spaces to extract information from different aspects. 
Specifically, the $i$-th layer user representation $\mathbf{h}_u^i$ is mapped into $H$ sub-spaces and the $h$-th sub-space is represented as:
 \begin{equation}
\label{sub_space_u}
\begin{aligned}
\mathbf{m}_{u}^{i,h}=\mathbf{W}_{u}^{i, h}\mathbf{h}_u^i+\mathbf{b}_{u}^{i, h}, \ \ \ \ \ \ \ h=1,2,\dots,H
\end{aligned}
\end{equation}
where $\mathbf{W}_{u}^{i, h}\in \mathbb{R}^{d_i \times p}$ and $\mathbf{b}_{u}^{i, h}\in \mathbb{R}^{p}$ are the transform matrix and bias vector for the $h$-head sub-space in the $i$-th user layer respectively, $p$ is the dimension of sub-space.
$\mathbf{m}_{u}^{i,h} \in \mathbb{R}^{p}$ is the $i$-th layer user latent representation under the $h$-th sub-space. It is noteworthy that, the $H$ head transformation can be concatenated for parallel computing and the result is $\mathbf{M}_{u}^{i}\in \mathbb{R}^{pH}$.

Similarly, the last layer item latent representation under the $h$-th sub-space can be represented as:
 \begin{equation}
\label{sub_space_i}
\begin{aligned}
\mathbf{m}_{v}^{L,h}=\mathbf{W}_{v}^{L, h}\mathbf{h}_v^L+\mathbf{b}_{v}^{L, h}, \ \ \ \ \ \ \ h=1,2,\dots,H
\end{aligned}
\end{equation}
and the concatenated result is presented as $\mathbf{M}_{v}^{L}\in \mathbb{R}^{pH}$.
\\\textbf{Interaction.} After extracting  informative knowledge from different sub-space, FE-Block applies fine-grained feature interactions to compute the relevance score $S_{u,v}^i$ between the $i$-th layer user latent representation and the last layer item latent representation. The relevance score $S_{u,v}^i$ is conducted as a sum of maximum similarity, calculated as follows:

\begin{equation}
\label{sum_max}
\begin{aligned}
S_{u,v}^{i} = \sum_{h_u = 1}^{H} \max \limits _{h_v \in 1,2,\dots, H}  \{  (\mathbf{m}_{u}^{i,h_u})^T \mathbf{m}_{v}^{L,h_v}\}.
\end{aligned}
\end{equation}
Note that the interaction mechanism in this step does not have any extra parameters, which is suitable for online serving. Finally, we calculate and sum the relevance score for each user layer to capture comprehensive multi-layer interactive signals: 
 \begin{equation}
\label{fetower_y}
\begin{aligned}
\hat{y} = \sum_{i=1}^{L}  S_{u,v}^{i}.
\end{aligned}
\end{equation}


In short, FE-Block module implements fine-grained and early user-item feature interaction explicitly, which improves the expression capability of the two-tower model significantly. More importantly, the ``user-item decoupling architecture'' paradigm can be maintained, ensuring high inference efficiency. The multi-head item latent representation $\mathbf{M}_{v}^{L}$ can be pre-stored for online serving. 
Besides, FE-Block replaces the original cosine similarity scoring approach (E.q.(\ref{inner_product})) with fine-grained sum of maximum cosine similarity (E.q.(\ref{sum_max}-\ref{fetower_y})), which is parameter-free and easy to deploy in the pre-ranking system.


\subsection{Contrastive Interaction Regularization}
FE-Block enhances the feature interactions of two-tower model via an explicit approach. In this subsection, we propose a contrastive interaction regularization module by applying contrastive learning to implicitly strengthen the information interaction between the two towers. In FE-Block, the multi-head latent representation $\mathbf{M}_{u}^{L}$ and $\mathbf{M}_{v}^{L}$ are the final user and item representation, respectively. To further strengthen interaction regularization  modeling, we introduce self-supervised signals to shorten the distance between user and positive items.

Hence, we leverage the positive instances to construct positive pairs $(\mathbf{M}_{u}^{L}, \mathbf{M}_{v}^{L})$, and $(\mathbf{M}_{u}^{L}, \mathbf{M}_{v'}^{L})$ are negative pairs when $v\ne v'$.
Following previous work~\cite{SimCLR}, we employ InfoNCE~\cite{infonce} to maximize the consistency of positive pairs and minimize the consistency of negative pairs:
\begin{equation}
\label{ssl}
\begin{aligned}
\mathcal{L}_{cir}= -\frac{1}{Q}\sum_{(u,v)\in Q}log\frac{exp(sim(\mathbf{M}_{u}^{L}, \mathbf{M}_{v}^{L})/ \tau )}{\sum\limits_{(u',v')\in N}exp(sim(\mathbf{M}_{u}^{L}, \mathbf{M}_{v'}^{L})/ \tau )},
\end{aligned}
\end{equation}
where $\tau$ is a temperature factor and $sim(\cdot)$ measures the similarity between two vectors, which is calculated by $sim(\mathbf{M}_{u}^{L}, \mathbf{M}_{v}^{L}) = (\mathbf{M}_{u}^{L})^T \mathbf{M}_{v}^{L}/(\|\mathbf{M}_{u}^{L}\| \cdot \|\mathbf{M}_{v}^{L}\|)$. $Q$ is the total number of positive  instances and $N$ is the total number of  samples.
By introducing contrastive interaction regularization, the information interaction between user and positive items can be effectively modeled.




\subsection{Training}
We use $Logloss$ to calculate the loss between the model prediction scores and the true labels, which is defined as follows:
 \begin{equation}
\label{e1}
\begin{aligned}
\mathcal{L}_{ctr} = -\frac{1}{N}\sum_{i=1}^{N}(y_{i}log(\hat{y_{i}}) + (1-y_{i})log(1-\hat{y_{i}}) ),
\end{aligned}
\end{equation}
where $y_{i}$ and $\hat{y_{i}}$ are the ground truth of $i$-th instance and score $\hat{y_{i}}$.

The ultimate purpose of the learning process is to minimize $\mathcal{L}_{ctr}$ and $\mathcal{L}_{cir}$, the final loss function is formulated as: 

 \begin{equation}
\label{e1}
\begin{aligned}
\mathcal{L} = \mathcal{L}_{ctr} + \lambda_{1}\mathcal{L}_{cir}+  \lambda_{2} \Vert \Theta \Vert_2,
\end{aligned}
\end{equation}
where $\lambda_{1}$, $\lambda_{2}$ are  hyper-parameters and $\Theta$ is the model parameters.






\section{Experiments}

In this section, we describe experiments in detail, including datasets, evaluation metrics, comparisons with SOTA baseline models and corresponding analyses. The experiment results on three public large-scale datasets demonstrate the effectiveness of IntTower on the task of CTR prediction. The code of IntTower is publicly available$^1$.

\subsection{Experimental Setting}
\subsubsection{Datasets}
We use three public available real-world datasets, whose statistics are summarized in Table~\ref{Table:stat.}. According to previous work~\cite{autoint,huang2019fibinet}, we split all samples randomly into two parts: 80\% for training and the rest for test.


\begin{table}[!t]

\footnotesize

\caption{\small{Statistics of datasets}}
 \resizebox{.47\textwidth}{!}{
\begin{tabular}{@{}cccccc@{}}
\toprule
Dataset         & Users       & Items  & User Field & Item Field & Samples    \\ \midrule
MovieLens-1M  & 6,040      & 3,952  & 5        & 3     & 1,000,000    \\
Amazon(Electro)      & 192,403      & 630,001  & 2      & 4     & 1,689,188  \\
Alibaba         & 1,061,768 & 785,597 & 9       & 6    & 26,557,961 \\ \bottomrule

\end{tabular}
}
\label{Table:stat.}
\vspace{-0.2cm}
\end{table}

\subsubsection{Evaluation Metrics}
Following previous work~\cite{din,autoint,huang2019fibinet}, we use three popular metrics to evaluate the performance . \textbf{Logloss} Logloss is a widely used metric in binary classification to measure the distance between two distributions. A lower bound of 0 for Logloss indicates that the two distributions are perfectly matched, and a smaller value indicates a better performance. \textbf{AUC} The area under the ROC curve (AUC) measures the probability that the model will assign a higher score to a randomly selected positive item than to a randomly selected negative item. \textbf{RelaImpr} RelaImpr metric~\cite{din} measures the relative improvement with respect to  models, which is defined as follows:

 \begin{equation}
\label{e1}
\begin{aligned}
RelaImpr = (\frac{AUC(measure\ model)-0.5}{AUC(base\ model)-0.5} -1)\times 100 \%
\end{aligned}
\end{equation}


\begin{table*}[htbp]
\normalsize
\setlength\tabcolsep{5pt}
\caption{\small{Performance comparison of different models. The base model of RelaImpr is Two-Tower, which is a popular neural network-based pre-ranking model. 
Boldface denotes the highest score and underline indicates the best result of the pre-ranking baselines. $\star$ represents significance level $p$-value $<0.05$ of comparing IntTower with the best pre-ranking baselines.}}
\begin{threeparttable}
\begin{tabular}{@{}ccccccccccc@{}}
\toprule
\multirow{2}{*}{Stage}  & \multirow{2}{*}{Model} & \multicolumn{3}{c}{MovieLens} & \multicolumn{3}{c}{Amazon} & \multicolumn{3}{c}{Alibaba} \\ \cmidrule(l){3-11} 
                              &                        & AUC     & Logloss  & RelaImpr & AUC       & Logloss    & RelaImpr   & AUC    & Logloss & RelaImpr \\ 
\cmidrule(l){1-11} 
\multirow{4}{*}{Pre-Ranking}                 & LR                     & 0.8501  & 0.4845   & -5.30\%  & 0.7952    & 0.4677     & -14.9\%    & 0.6496 & 0.2350  & -5.85\%  \\  & Two-Tower                   & 0.8697  & 0.4559   & 0\%      & 0.8469    & 0.4313     & 0\%        & 0.6579 & 0.2292  & 0\%      \\
                              & DAT                    & 0.8712  & 0.4556   & 0.40\%   & 0.8480    & 0.4278     & 0.31\%     & 0.6598 & 0.2279  & 0.56\%   \\
                              & COLD                & \underline{0.8836}  & \underline{0.3297}   & 3.75\%   & \underline{0.8633}    & \underline{0.3402}     & 4.72\%     & \underline{0.6816} & \underline{0.2248}  & 14.28\%  \\ \midrule
\multirow{4}{*}{Ranking$^*$}   & Wide\&Deep                   & 0.8820  & 0.3344   & 3.32\%   & 0.8615    & 0.3409     & 4.20\%     & 0.6814 & 0.2250  & 14.15\%  \\
                              
                              & DeepFM            & 0.8920  & 0.3211   & 6.03\%   & 0.8643    & 0.3405     & 5.05\%     & 0.6820 & 0.2247  & 14.53\%  \\
                              & DCN                    & 0.8964  & 0.3151   & 7.22\%   & 0.8665    & 0.3366     & 5.65\%     & 0.6831 & 0.2244  & 15.22\%  \\
                              & AutoInt                & 0.8948  & 0.3192   & 6.79\%   & 0.8686    & 0.3351     & 6.25\%     & \textbf{0.6867} & \textbf{0.2238}  & \textbf{17.49\%}  \\ \midrule
                              & IntTower (ours)               & \textbf{0.8974}$^\star$  & \textbf{0.3128}$^\star$   & \textbf{7.49\%}   & \textbf{0.8696}$^\star$    & \textbf{0.3309}$^\star$     & \textbf{7.91\%}     & 0.6827$^\star$ & 0.2245$^\star$  & 14.97\%  \\ \bottomrule
\end{tabular}
\begin{tablenotes}    
\huge \item[*] \small \textbf{Generally speaking, the prediction accuracy of ranking models is much higher than that of the pre-ranking models by a large margin~\cite{cold,fscd}.
To fully demonstrate the superior prediction accuracy of IntTower, we compare it with some SOTA ranking models and are delighted to find that the performance is \textit{comparable} and even \textit{exceeded} in some cases.
}
\end{tablenotes} 
\end{threeparttable}
\label{performance}
\end{table*}

\subsubsection{Competing Models.}
We compare IntTower with the following models, and for fairness, we classify the compared models into two classes. (A) Pre-Ranking Models. (B) Ranking Models. 
The pre-ranking models include \textbf{LR}~\cite{mcmahan2013ad}, \textbf{Two-Tower}~\cite{dssm}, \textbf{DAT}~\cite{dat}, \textbf{COLD}~\cite{cold}, and  the ranking models include \textbf{Wide\&Deep}~\cite{widedeep}, \textbf{DeepFM}~\cite{guo2017deepfm}, \textbf{DCN}~\cite{dcn}, \textbf{AutoInt}~\cite{autoint}.


\subsubsection{Implementation Details}
We implement the proposed IntTower and other compared baselines with Pytorch~\cite{NEURIPS2019_9015} and empirically set the feature embedding dimension $d$ to 32, and the batch size to 2048. Besides, for all deep models, we set the number of hidden layers $L$ as 3 and the number of hidden units as $[300,300,128]$.
For IntTower, we use three FE-Blocks by default, where the number of sub-spaces is equal to the number of features in the user tower, \textit{i.e.}, $H=m$ and the dimension of each sub-space (head size) is set to 64. All models are trained by the Adam~\cite{https://doi.org/10.48550/arxiv.1412.6980} optimizer with the initial learning rate setting to 0.001. Batch Normalization~\cite{bn} and Dropout~\cite{dropout} is applied to avoid overfitting. To ensure a fair comparison, other hyper-parameters such as training epochs are adjusted individually for all models to obtain the best results. We conduct experiments with 2 Tesla 3090 GPUs.


\subsection{Result of Model Accuracy}
We compare the overall performance with some SOTA pre-ranking and ranking models, results are summarized in Table~\ref{performance}. Generally speaking, the prediction accuracy of ranking models is much higher than that of the pre-ranking models by a large margin~\cite{cold,fscd}.
To fully demonstrate the superior prediction accuracy of IntTower, we compare it with some ranking models and are delighted to find that the performance is \textit{comparable} and even \textit{exceeded} in some cases. 
Noted that, for the concern of fairness, we repeat each experiment 10 times to obtain the best results. From which, we obtain the following observations: (1) Among the pre-ranking models, LR achieves the worst performance in comparison with the neural network-based models. Two-Tower leverages the FC layers to capture the feature interactions within the user/item towers independently. Though DAT implicitly models the information interaction between the two towers, the performance gains are limited.
COLD with single-tower structure outperforms other pre-ranking models while the increased inference burden is also noticeable. (2) Compared with the pre-ranking models, ranking models (\textit{e.g.}, DeepFM and AutoInt) achieve significant improvement by a large margin. Ranking models with single-tower structure attribute to capturing higher-order feature interactions between user and item, which is vital for CTR prediction.
  However, these models cannot be deployed in the pre-ranking system without massive engineering optimization. Therefore, they are more suitable for serving in the ranking stage with fewer candidates. (3) IntTower outperforms SOTA pre-ranking models by a large margin and even some ranking models (\textit{e.g.}, Wide\& Deep and DeepFM), indicating superior prediction performance. Even in the face of industrial SOTA ranking models DCN and AutoInt, IntTower still achieves a comparable performance. We attribute this great improvement to the effective interaction modeling between user and item towers from the explicit and implicit perspectives.
  Moreover, IntTower keeps the ``user-item decoupling architecture'' paradigm, which is highly efficient 
  , making our model suitable for the pre-ranking system.


\subsection{Result of Model Efficiency}
\label{sec:efficient}
\subsubsection{Training Efficiency and Model Size}
For CTR prediction, the model training efficiency is crucial as it will affect the model update frequency and further affect the prediction accuracy. 
In this part, we calculate the parameters and running time of different models over the MovieLens and Amazon, shown in Table~\ref{runtime}.

Not surprisingly, LR has the smallest parameters and lowest training time due to its shallow structure. 
Compared with the Two-Tower, the increased parameters and training time of IntTower is negligible. Besides, the single-tower structure models have more parameters and higher training time than IntTower.
IntTower employs Light-SE feature attention mechanism and the FE-Block introduces only a few parameters to model fine-grained feature interaction.


\begin{table}[!t]
\caption{\small{Training efficiency comparison of different models in terms of Model Size and Running Time in an epoch. For the model with two-tower structure , we only calculate the parameters of the user tower since only the user tower is required for online inference.}}
\resizebox{0.8\linewidth}{!}{
\begin{tabular}{@{}ccccc@{}}
\toprule
               & \multicolumn{2}{c}{MovieLens} & \multicolumn{2}{c}{Amazon} \\ \midrule
Model & Params      & Run Time        & Params        & Run Time       \\ \midrule
LR             &   5.24$\times 10^4$          &    11s             &   5.10$\times 10^5$         &   27s            \\
Two-Tower             &    5.03$\times 10^5$         &     15s            &    6.26$\times 10^6$        &   35s            \\
COLD            &   6.83$\times 10^5$          &  18s               & 8.58$\times 10^6$          &   42s            \\
 DCN              &  6.84$\times 10^5$           &    19s             &     8.58$\times 10^6$       &      43s        \\
 AutoInt             &    6.95$\times 10^5$         &  19s               &      8.59$\times 10^6$         &     42s            \\ 
IntTower             &    5.67$\times 10^5$         &  16s               &       6.29$\times 10^6$         &     40s            \\
 \bottomrule
 
\end{tabular}}
\label{runtime}
\vspace{-0.2cm}
\end{table}


\begin{figure}[htbp]
\centering
\includegraphics[scale=0.32]{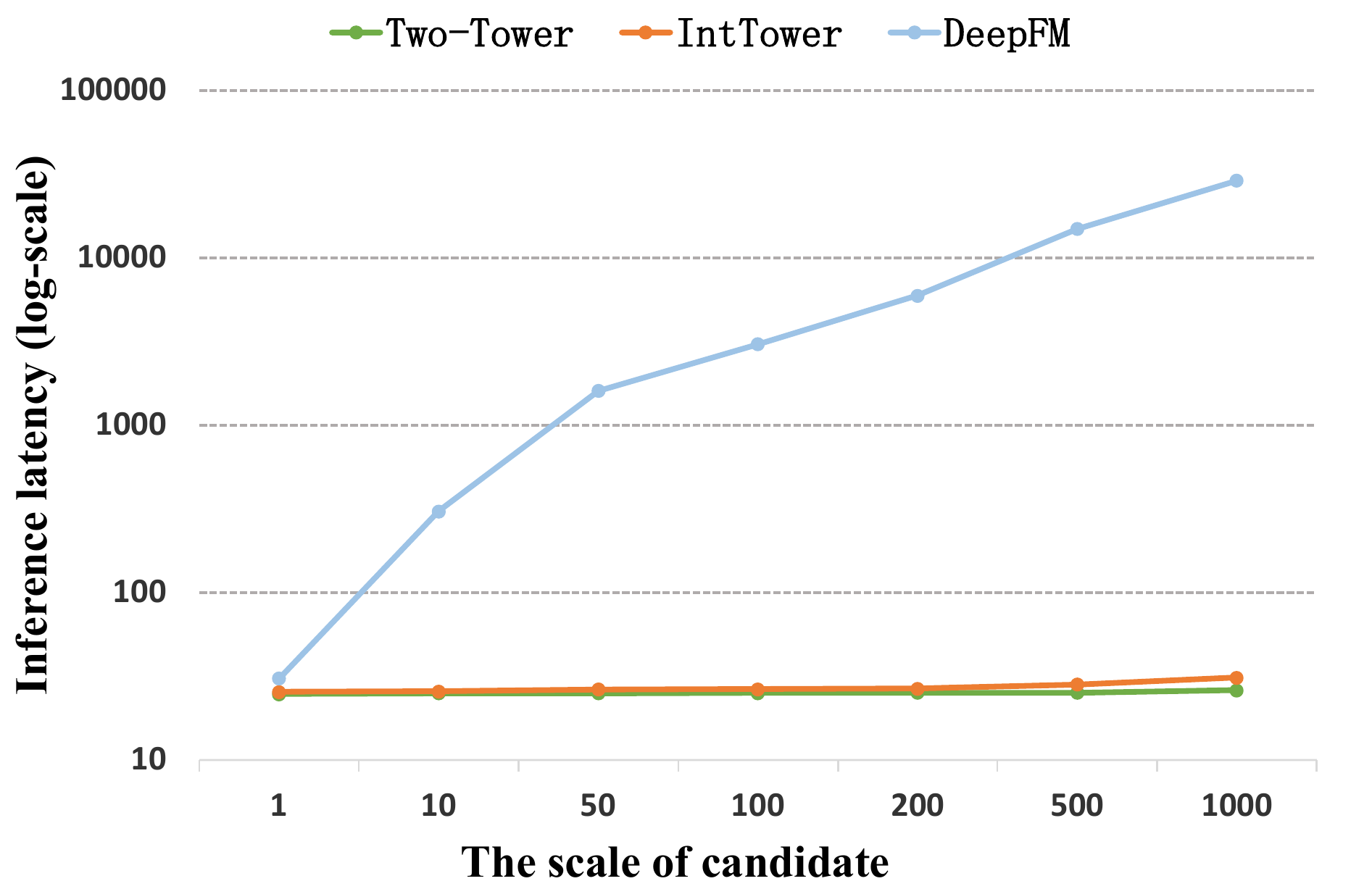}
\caption{\small{Inference latency for different candidate set sizes. We use MovieLens, the number of users is 1$\times 10^{5}$, test on a Tesla 3090 Gpu.} }
\label{latency}
\vspace{-0.2cm}
\end{figure}

\subsubsection{Serving Efficiency}
In industrial multi-stages cascade systems, each stage is subject to strict latency limits, \textit{e.g.}, 10$\sim$20 milliseconds. In this case, pre-ranking system is typically designed as low-latency system because the size of the candidates is thousands for each user request, which is much more than that of the ranking stage. Therefore, high serving efficiency is indispensable for a pre-ranking model. In this subsection, we discuss the model serving efficiency.


In Figure~\ref{latency}, we show the inference latency of different models over different candidate sizes. For Two-Tower and IntTower, we store the obtained item representations into faiss~\cite{johnson2019billion}, and during inference, we retrieve the corresponding representations from faiss for prediction. The potential of IntTower can be clearly illustrated in Figure~\ref{latency}. Compared with the Two-Tower with the highest efficiency, the increased latency of IntTower is acceptable given different candidate scales. 
Benefiting from the ``user-item decoupling architecture'' paradigm, the time complexity of both Two-Tower and IntTower is $O(N+kM)$, where $k$ is the number of candidates, $N$ is the cost of a neural network inference and $M$ is the cost of a vector retrieval and score prediction.
Therefore, IntTower has superior online serving efficiency and is suitable for pre-ranking systems. 
Instead, for the single-tower model (\textit{e.g.}, DeepFM), each user-item pair require a neural network inference, resulting in $O(kN)$ complexity ($N\gg M$).


\subsection{Ablation Study}
                

\begin{figure}[htbp]
	\centering
	\setlength{\belowcaptionskip}{-0.3cm}
	\setlength{\abovecaptionskip}{0cm}
	\subfigure[\small{MovieLens}]{
		\label{fig:ablation:ml} 
		\includegraphics[height=75pt]{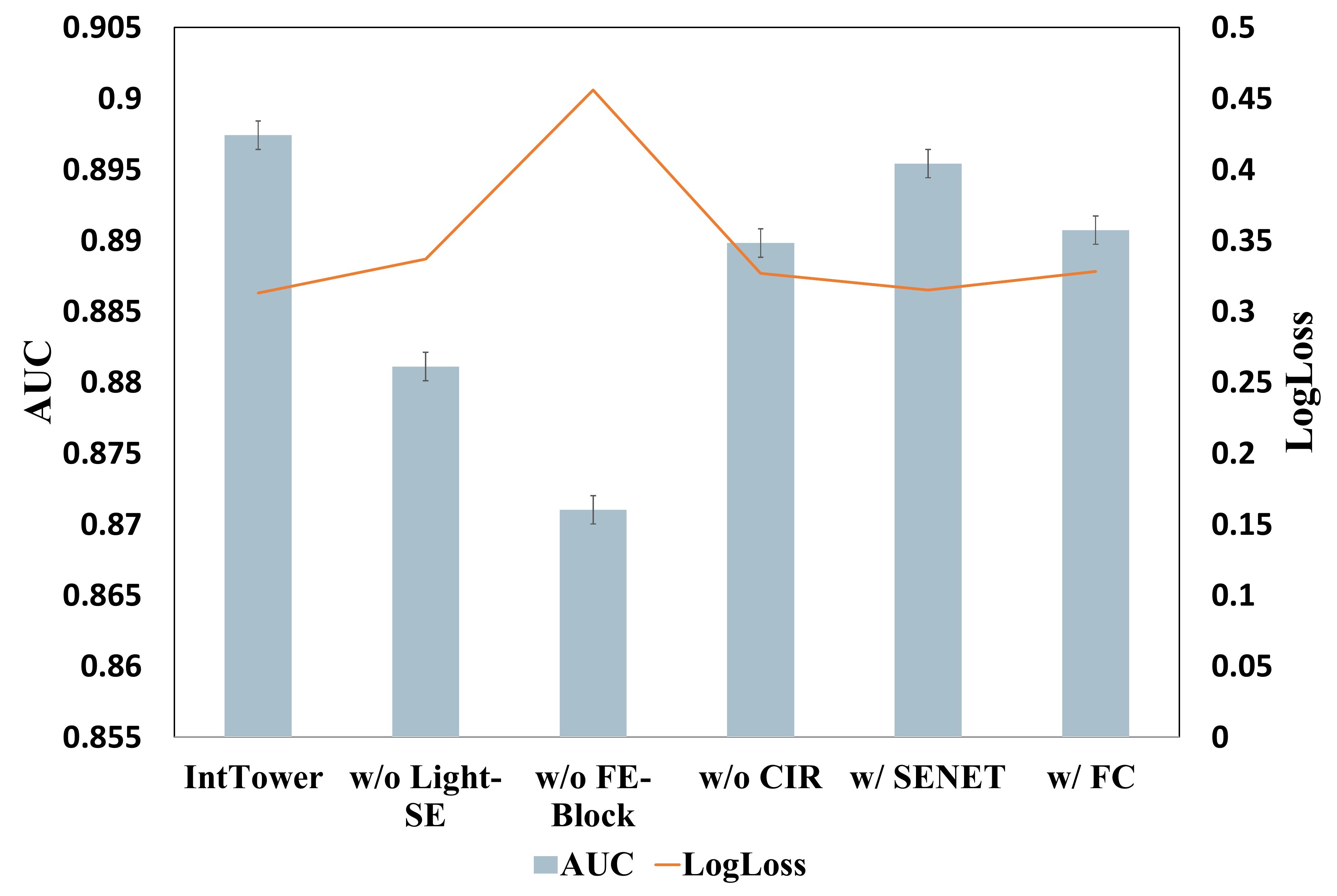}}
	\subfigure[\small{Amazon}]{
		\label{fig:ablation:amazon} 
		\includegraphics[height=75pt]{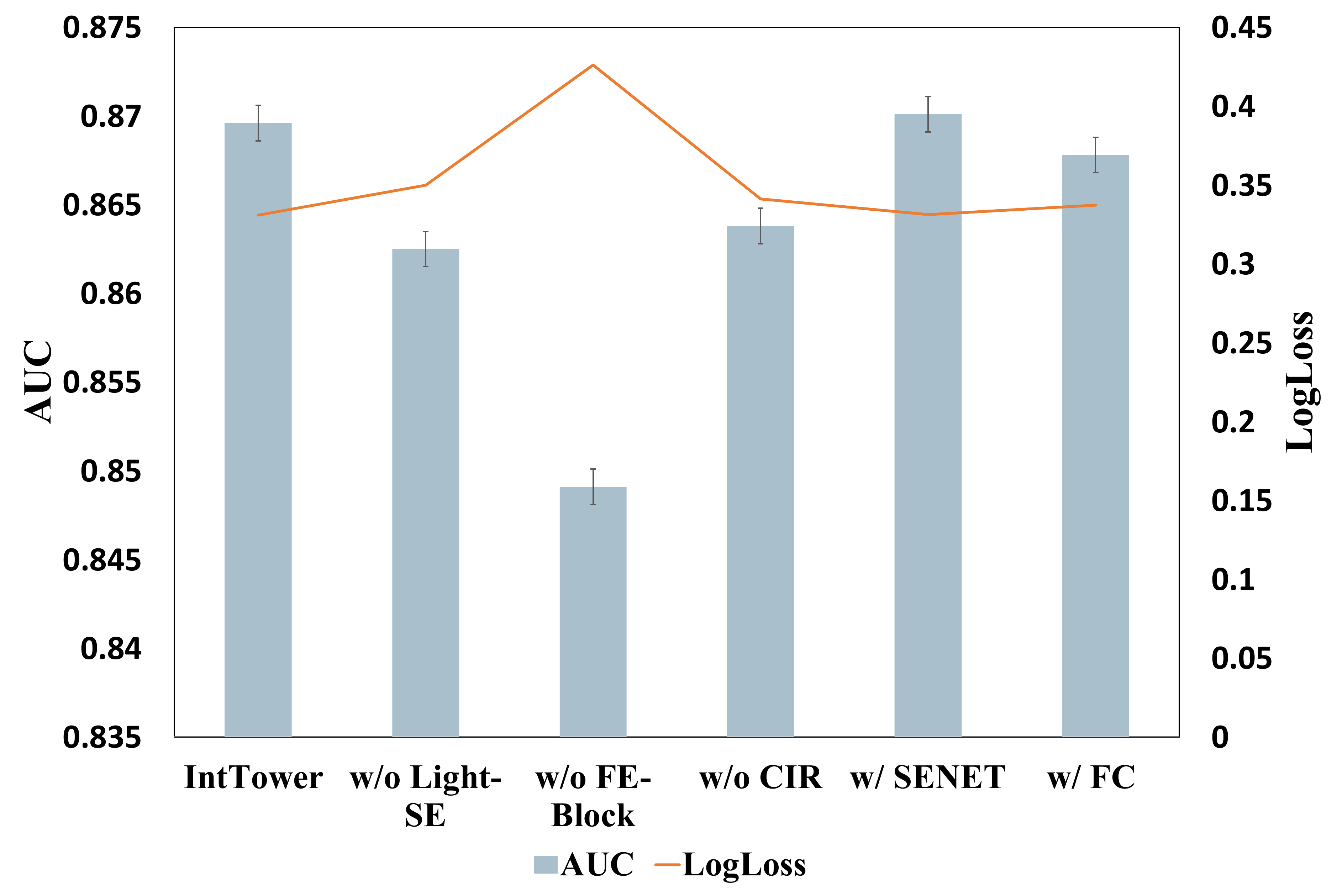}}
	\caption{\small{The impact of different components in IntTower.}}
	\vspace{-0.2cm}
	\label{fig:ablation}
\end{figure}


\subsubsection{The Impact of Different  Components}
In this section, we conduct ablation experiments on the IntTower to better understand the importance of different components. We first remove the three modules of IntTower; then we replace the Light-SE with SENET, and replace the FE-Block with a FC layer via $FC(\mathbf{h}_u^i||\mathbf{h}_v^L)$.
From Figure~\ref{fig:ablation}, we observe the following results: (1) Each component plays an active role in the performance of the IntTower, and the removal of any component  will result in a degradation of performance. (2) FE-Block is the most important part of IntTower. Removing it will lead to a huge performance drop, which indicates that the explicit feature interaction modeling between user and item towers is statistically significant for two-tower.
(3) After replacing Light-SE with SENET, the performance decreases on MovieLens and slightly improves AUC on Amazon. The above phenomena indicates that Light-SE has fully comparable performance to SENET but less parameters. 
(4) After replacing FE-Block with a FC layer, the performance decreases but still achieves great improvement compared with the two-tower model, demonstrating the importance of interaction modeling. However, FC layer is not friendly to online serving due to extra parameters. Instead, the \textit{Interaction} step in FE-Block is parameter-free, which is suitable for online deployment.

\begin{figure}[htbp]
	\centering
	\setlength{\belowcaptionskip}{-0.3cm}
	\setlength{\abovecaptionskip}{0cm}
	\subfigure[\small{MovieLens}]{
		\label{fig:layer:ml} 
		\includegraphics[height=75pt]{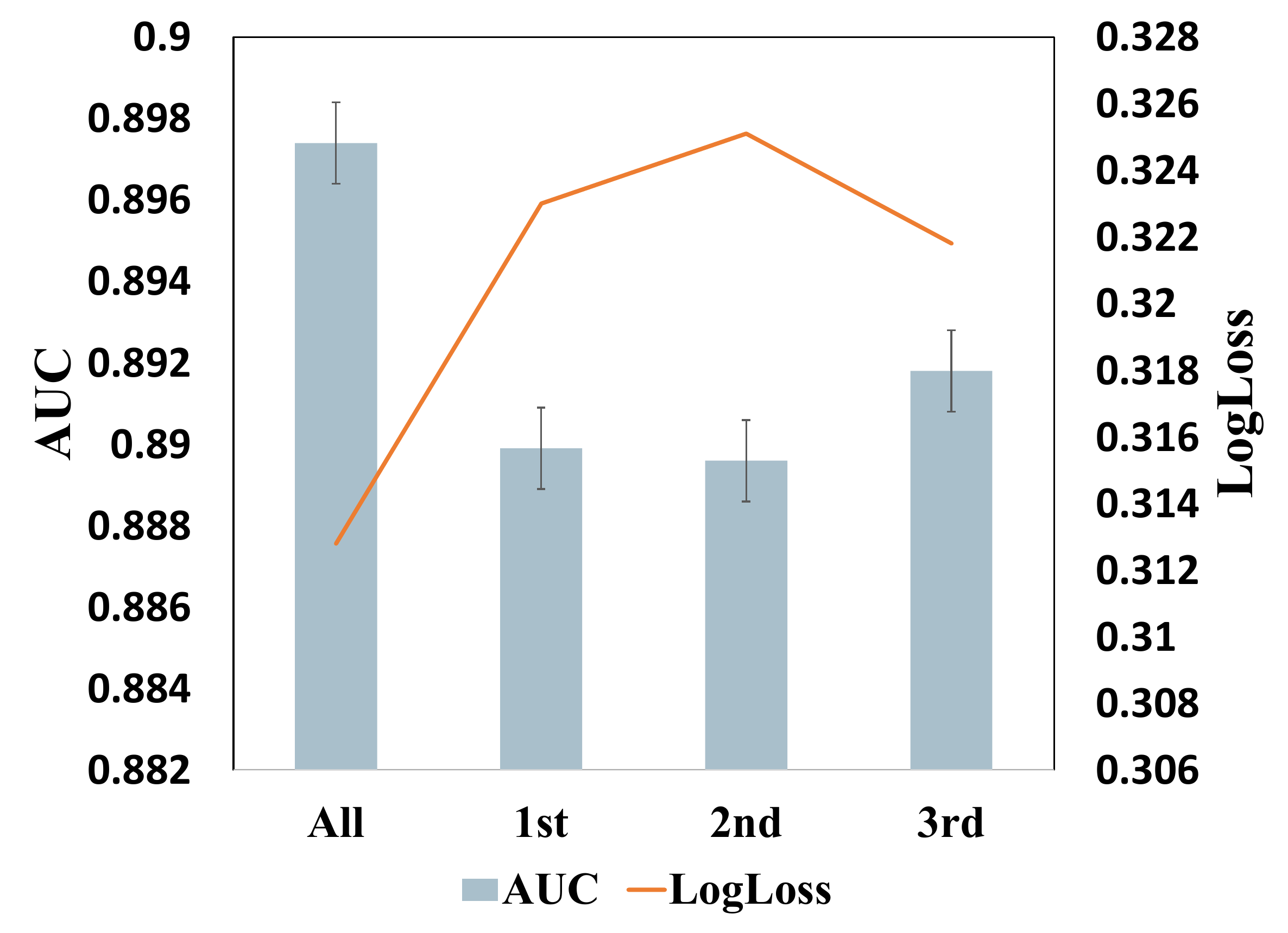}}
	\subfigure[\small{Amazon}]{
		\label{fig:layer:amazon} 
		\includegraphics[height=75pt]{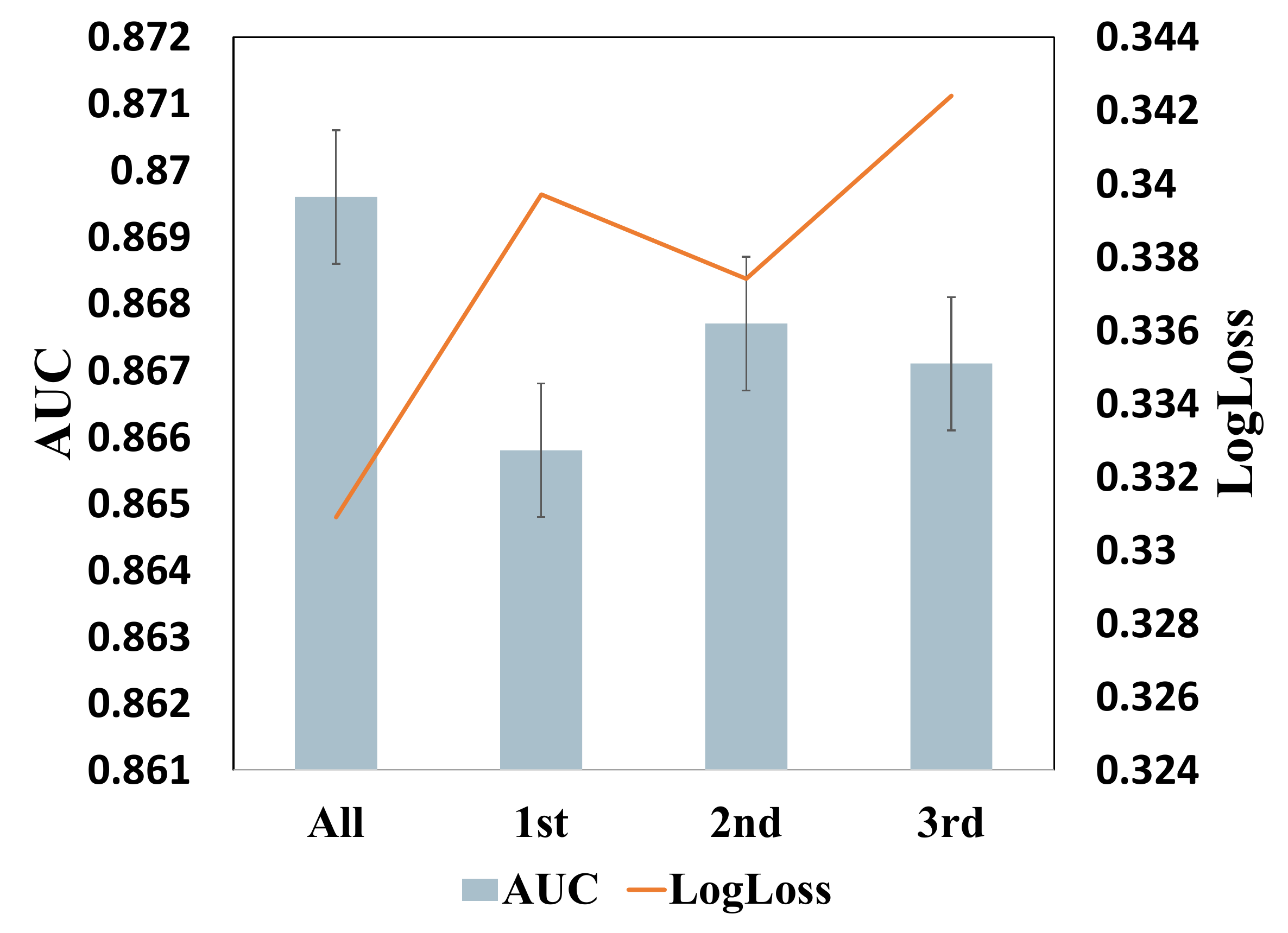}}
	\caption{\small{The performance of different user tower layers.}}
	\vspace{-0.2cm}
	\label{fig:layers}
\end{figure}

\subsubsection{The Impact of FE-Block in Different Layers} 
In IntTower, we deploy multiple FE-Blocks to capture interactive signals between multiple user layers $\{\mathbf{h}_u^1, \mathbf{h}_u^1, \dots, \mathbf{h}_u^L\}$ and the last item layer $\mathbf{h}_v^L$.
In order to further study the influence of the user layer in FE-Block, we perform experiments over MovieLens and Amazon datasets by using a single FE-Block and summarize the experimental results in Figure~\ref{fig:layers}.
We can find some observations as follows: (1) Deploying a single FE-Block still achieves significant performance, demonstrating the effectiveness of the fine-grained feature interaction modeling.
(2) Performing multiple FE-Blocks over multiple user layers can obtain significant improvement since different level feature interactions can be captured synthetically.
Remarkably, the effect of only one FE-Block is also quite well, so we can reduce the number of FE-Blocks to further reduce online latency.



\subsection{Hyper-parameter Study}

\subsubsection{The impact of the number of heads}
In order to investigate the effect of the number of heads, experiments are implemented over MovieLens and Amazon datasets and summarized in Table~\ref{head}. From the results we can get the following observations: 
(1) When increasing the number of heads, the overall performance shows a trend of increasing first and then decreasing, indicating that increasing a certain number of heads helps boost performance.
(2) For the MovieLens dataset, the optimal number of heads is achieved with a small number $\{2, 2\}$. We speculate the reason is that the model with more heads tend to overfit on the MovieLens dataset. Instead, the optimal number for Amazon dataset is much larger ($\{16, 16\}$).


\begin{table}[!tbp]
\scriptsize
\caption{\small{Performance \textit{w.r.t.} the number of heads in FE-Block.}}
\setlength\tabcolsep{4pt}
\resizebox{0.8\linewidth}{!}{
\begin{tabular}{@{}cccccc@{}}
\toprule
\multirow{2}{*}{User Head} & \multirow{2}{*}{Item Head} & \multicolumn{2}{c}{MovieLens} & \multicolumn{2}{c}{Amazon} \\ \cmidrule(l){3-6} 
                              &                               & AUC           & Logloss       & AUC              & Logloss          \\ \midrule
1                             & 1                             & 0.8957        & 0.3160        & 0.8669           & 0.3513           \\
1                             & 2                             & 0.8968        & 0.3144        & 0.8653           & 0.3397           \\
1                             & 4                             & \textbf{0.8994}        & \textbf{0.3110}        & 0.8671           & 0.3368           \\
1                             & 8                             & 0.8953        & 0.3215        & 0.8660           & 0.3366           \\
1                             & 16                            & 0.8987        & 0.3126        & 0.8689           & 0.3328           \\
1                             & 32                            & 0.8936        & 0.3184        & \textbf{0.8694}           & \textbf{0.3322}           \\ \midrule
2                             & 2                             & \textbf{0.9016}        & \textbf{0.3077}        & 0.8672           & 0.3356           \\
4                             & 4                             & 0.8944        & 0.3215        & 0.8668           & 0.3379           \\
8                             & 8                             & 0.8961        & 0.3158        & 0.8679           & 0.3344           \\
16                            & 16                            & 0.8920        & 0.3217        & \textbf{0.8705}           & \textbf{0.3307}           \\
32                            & 32                            & 0.8953        & 0.3166        & 0.8701           & 0.3309           \\ 
64                             & 64                             & 0.8974        & 0.3128        & 0.8696           & 0.3309           \\ \bottomrule
\end{tabular}}
\label{head}
\vspace{-0.36cm}
\end{table}
\subsubsection{The impact of the head sub-space dimension}
\begin{figure}[htbp]
	\centering
	\setlength{\belowcaptionskip}{-0.3cm}
	\setlength{\abovecaptionskip}{0cm}
	\subfigure[\small{MovieLens}]{
		\label{fig:head_size:ml} 
		\includegraphics[height=80pt]{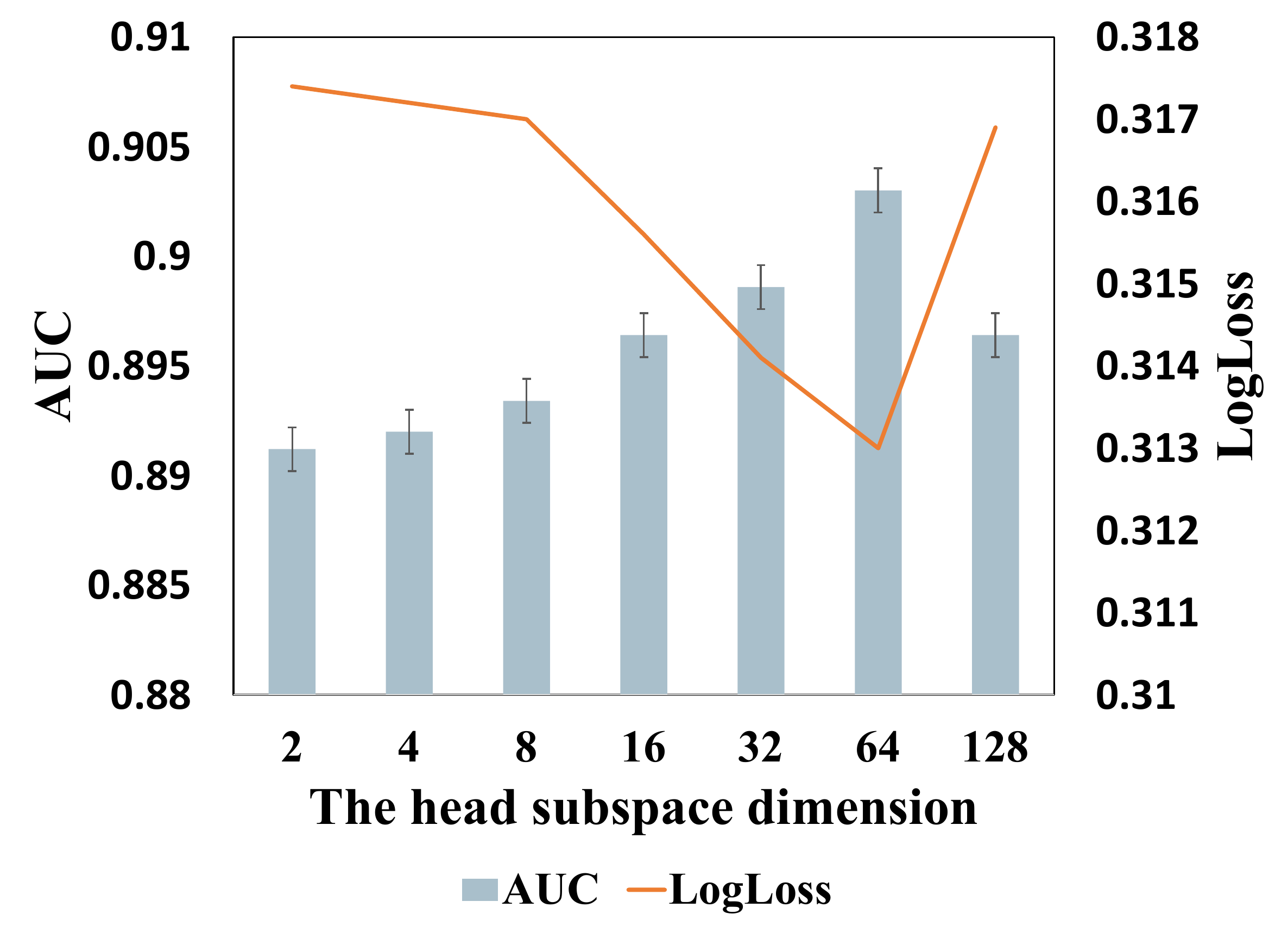}}
	\subfigure[\small{Amazon}]{
		\label{fig:head_size:amazon} 
		\includegraphics[height=80pt]{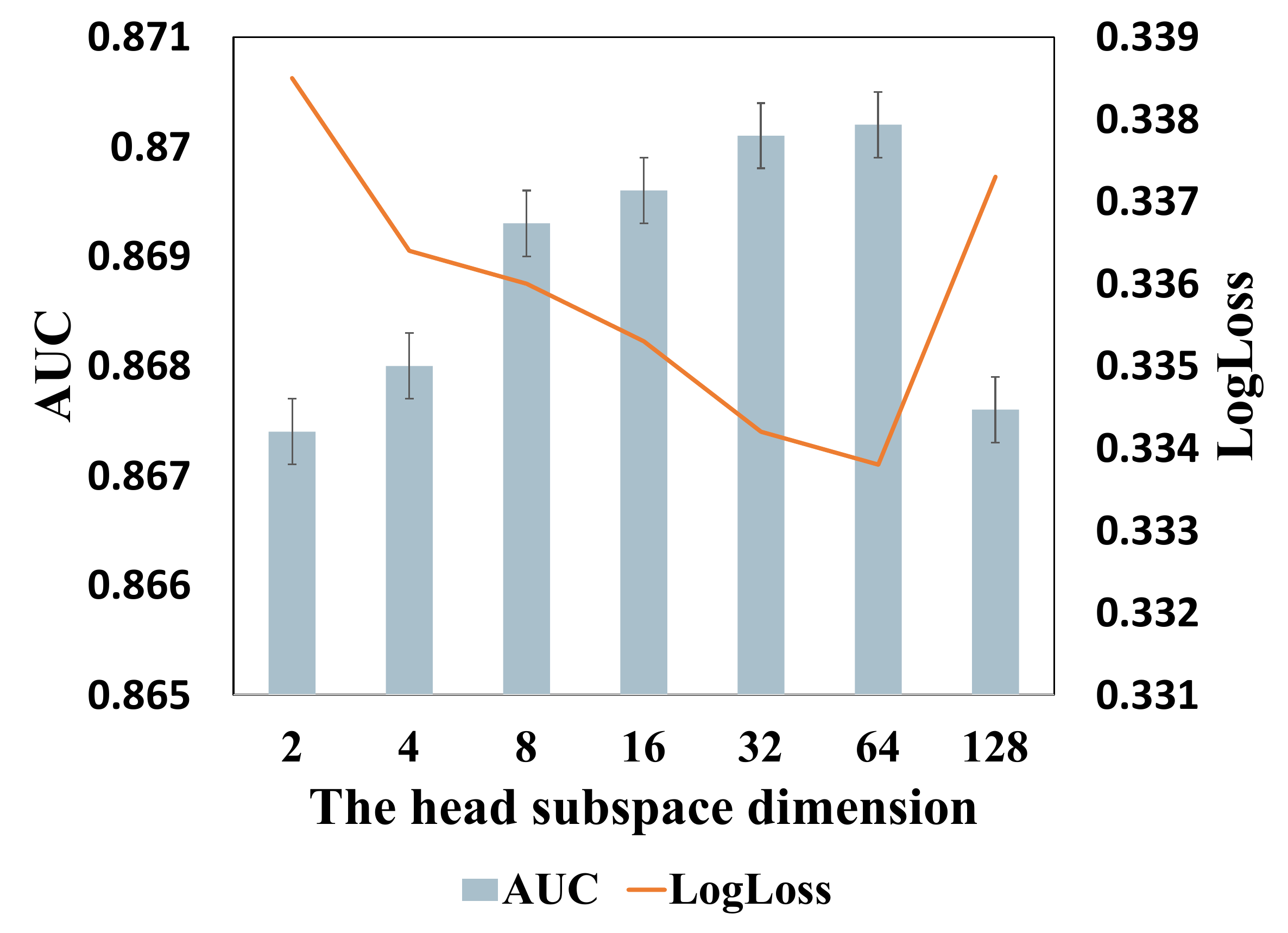}}
	\caption{\small{Performance \textit{w.r.t.} the head subspace dimension .}}
	\vspace{-0.4cm}
	\label{head_size}
\end{figure}

Next, we investigate the impact of the sub-space dimensions in FE-Block and the results are depicted in Figure~\ref{head_size}. We can observe that, as the sub-space dimension increases, the model performance increases first and then decreases. 
The optimal performance can be achieved when the sub-space dimension reaches 64.
The reason is that enlarging the dimension attributes to capture informative interactions due to more parameters. However, more parameters also tend to overfit, thus degrading the performance.





\subsection{Visualization of IntTower}



\begin{figure}[!t]
\centering
\subfigure[Two-Tower]{               
\includegraphics[scale=0.19]{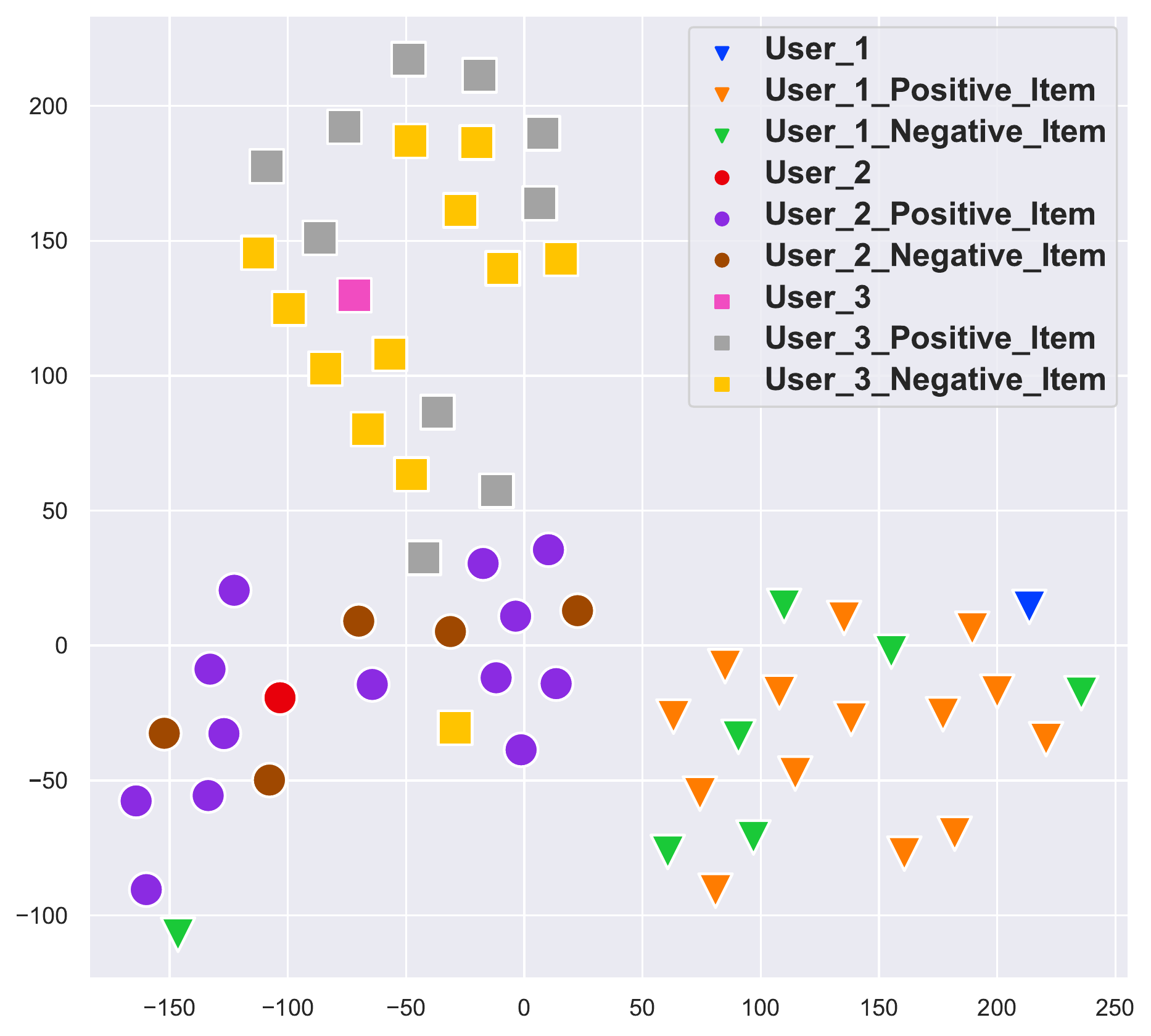}}
\hspace{0in}
\subfigure[IntTower]{
\includegraphics[scale=0.19]{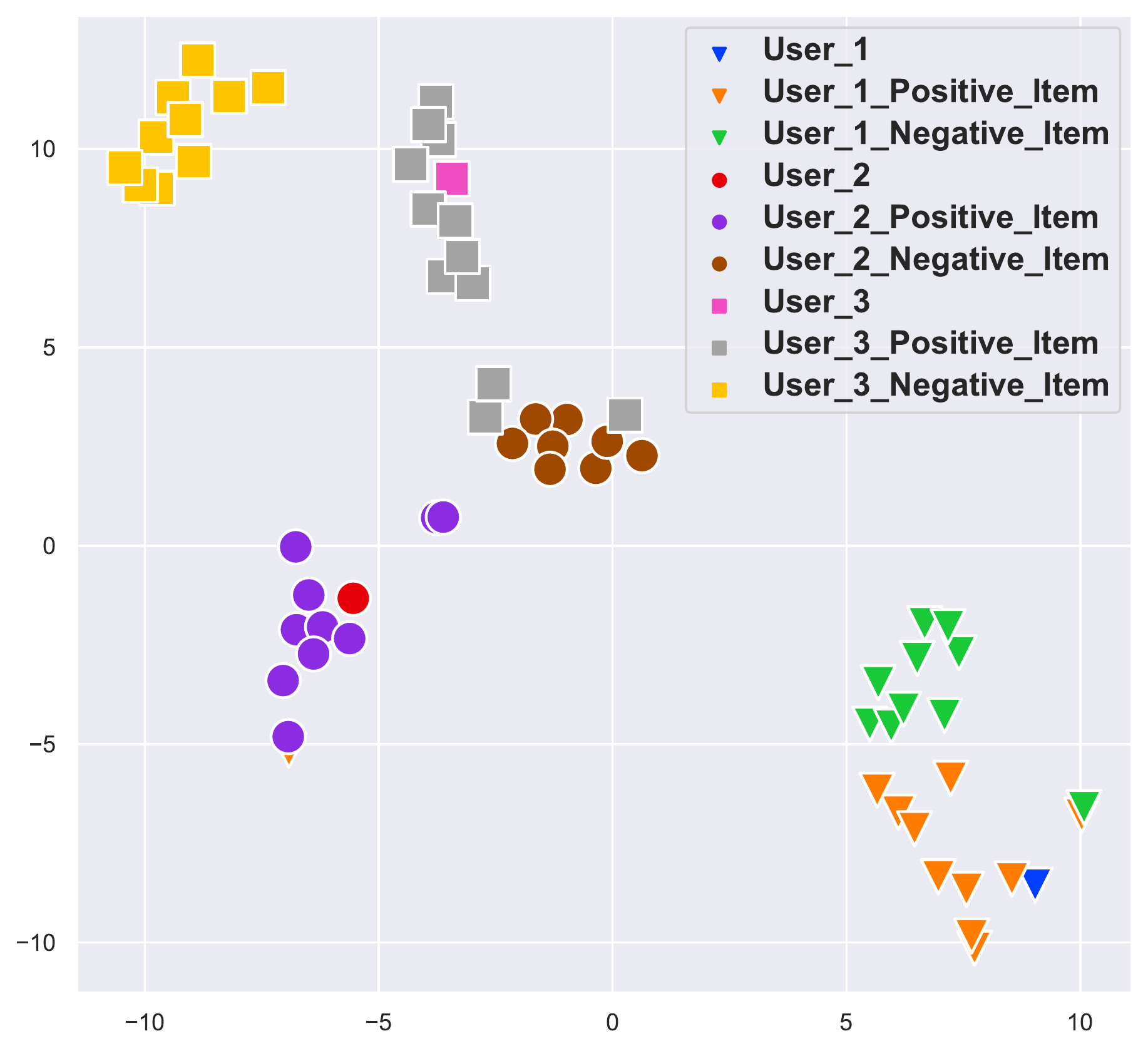}}
\caption{\small{Visualization of user and item representations. Points with the same shape represent a user and the corresponding items.}}
\label{tsne}
\vspace{-0.2cm}
\end{figure}

To deeply investigate the properties of original Two-Tower and IntTower, we visualize the user and item representations by projecting into a two-dimensional space with t-SNE~\cite{t-SNE}. 
Points with the same shape correspond to a user and the corresponding positive and negative items in Amazon Dataset. As shown in Figure~\ref{tsne}, for Two-Tower, the positive and negative items are around the user and mixed together, which is difficult to distinguish.
By contrast, for IntTower, user and positive items are almost in the same cluster, while negative items are far away from the user. 
This phenomenon demonstrates that IntTower can better infer users' preferences and distinguish between positive and negative items.

\section{Application: Advertisement Pre-Ranking System}
In this section, we deploy IntTower in a large-scale industrial pre-ranking system to validate its effectiveness.

\subsection{System Overview}
\begin{figure}[!t]
\centering
\includegraphics[scale=0.45]{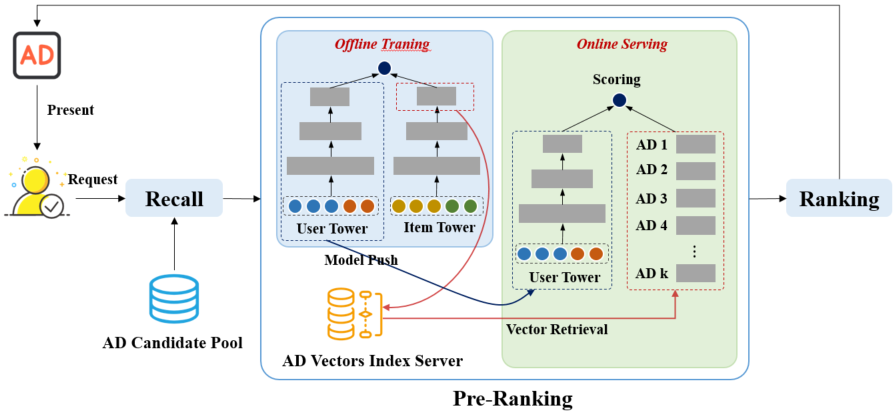}
\caption{\small{Overview of pre-ranking model in  advertisement system.}}
\label{system}
\vspace{-0.2cm}
\end{figure}

The overview of pre-ranking system in a large-scale advertisement platform is presented in Figure~\ref{system}.
When a user request arrives, it triggers the recall, pre-ranking, and ranking recommendation pipeline. Pre-ranking system scores the candidate ads retrieved from the recall stage and generates top-$k$ high-quality ads for the subsequent ranking stage.
Specifically, pre-ranking system consists of two core modules: \textit{Offline Training} and \textit{Online Serving}. 

Offline training is periodically performed over the latest user behavior logs to update the model parameters. 
Take the two-tower model as an example, features are divided into user-related $x_{user}$ and ad-related $x_{ad}$, which are fed into user tower and ad tower for training, respectively. 
Upon the finish of model training, the user tower model is deployed on the online server for prediction while the ad tower model is utilized to infer ads representation vectors $\mathbf{e}_{ad}$, which will be stored in the AD Vectors Index Server  for the online serving.
During the online serving phase, a user request will trigger user and contextual feature gathering, and these features will be concatenated into a user feature vector $x_{user}$. Then, feature vector $x_{user}$ is fed into the user tower model for inferring user representation $\mathbf{e}_{user}$. A candidate ad ID list returned from the recall stage is used to retrieve a group of ad representations $\{\mathbf{e}_{ad\ 1}, \mathbf{e}_{ad\ 2}, \dots, \mathbf{e}_{ad\ k}\}$ from the AD Vectors Index Server. Finally, the online scoring server performs $k$ prediction to obtain predicted scores $Pr(y_i|\mathbf{e}_{user}, \mathbf{e}_{ad\ i})$ for each candidate ad $i$, and a sorted ad ID list is organized according to the pre-defined ranking function (\textit{e.g.}, eCPM) before sending to the ranking stage.

\subsection{Dataset and Compared Models}
We collect and sample 8 consecutive days of user behavior records from a large-scale advertisement platform, where millions of daily active users interact with advertisements and tens of millions of user logs are generated. 
The user and contextual related features $x_{user}$ include user profile features (\textit{e.g.}, gender), user behavior features (\textit{e.g.}, list of ads clicked by users), user statistics features (\textit{e.g.}, number of ads clicked by users), as well as contextual features (\textit{e.g.}, time). 
Additionally, the ad related features $x_{ad}$ include ad task features (\textit{e.g.}, task id) and ad statistics features (\textit{e.g.}, number of clicks on the ad). For the categorical features, the embeddings are obtained by embedding look-up, while the numerical feature embeddings are learned via the AutoDis~\cite{autodis}.

We compare IntTower with  pre-ranking and ranking models, respectively. For the pre-ranking model, we choose a highly-optimized DNN-based Two-Tower model. For the ranking model, we choose DeepFM~\cite{guo2017deepfm} and EDCN~\cite{edcn}, which are deployed in the large-scale advertisement business system.


\subsection{Performance Comparison}
The performance comparison on the large-scale advertising dataset is presented in Table~\ref{table:huawei}, where we can observe that  IntTower outperforms Two-Tower model by a significant margin in terms of AUC and Logloss, even surpasses ranking model DeepFM and achieving a performance close to EDCN.
By enhancing the information interaction between user and item towers, IntTower achieves significant performance improvement compared with the original Two-Tower model.
Moreover, as presented in Figure~\ref{latency}, the inference latency of IntTower is several orders of magnitude lower than the single-tower structure models.
Therefore, IntTower has comparable prediction accuracy while higher inference efficiency than the ranking models, which fits perfectly with industrial pre-ranking systems.

\begin{table}[!t]
\setlength\tabcolsep{6pt}
\footnotesize
\caption{\small{The performance comparison in a  advertisement system.}}
\begin{tabular}{@{}ccccc@{}}
\toprule
Stage & Model               & AUC                        & Logloss & RelaImpr                   \\ \midrule
\multirow{2}{*}{Pre-Ranking}&Two-Tower  &  0.7550 &   0.0536    &  0\%  \\
&IntTower & 0.7603   &  0.0523 &  2.08\% \\
\midrule
\multirow{2}{*}{Ranking}&DeepFM& 0.7591 & 0.0525      &  1.61\%        \\
&EDCN & 0.7612  & 0.0518 &  2.43\% \\
 \bottomrule
\end{tabular}
\label{table:huawei}
\vspace{-0.2cm}
\end{table}

\section{Conclusion}
In this paper, we propose the Interaction enhanced Two-Tower model (IntTower) to boost the performance of two-tower model by capturing information interaction between user and item towers and preserving high inference efficiency with ``user-item decoupling architecture'' paradigm. The IntTower includes Light-SE, FE-Block and  CIR modules. Specifically, Light-SE module can identify the importance of different features and obtain refined feature representations. FE-Block module performs fine-grained and early feature interactions to capture the interactive signals between user and item towers explicitly while CIR module leverages a contrastive interaction regularization to further enhance the interactions implicitly. 
Comprehensive experiments show that IntTower outperforms the SOTA pre-ranking models significantly. Moreover, we further verify the effectiveness of IntTower on a large-scale advertisement pre-ranking system.

\newpage
\normalem
\balance
\bibliographystyle{ACM-Reference-Format}
\bibliography{sample-base}

\end{document}